\documentclass{article}

\usepackage{arxiv}

\usepackage[utf8]{inputenc} % allow utf-8 input
\usepackage[T1]{fontenc}    % use 8-bit T1 fonts
\usepackage{url}            % simple URL typesetting
\usepackage{booktabs}       % professional-quality tables
\usepackage{amsfonts}       % blackboard math symbols
\usepackage{nicefrac}       % compact symbols for 1/2, etc.
\usepackage{microtype}      % microtypography
\usepackage{lipsum}		% Can be removed after putting your text content
\usepackage{graphicx}
\usepackage[square,sort,comma,numbers]{natbib}
\usepackage{doi}

\usepackage{blindtext}
\usepackage{multicol}
\usepackage{floatpag}

\usepackage{float}
\usepackage[T1]{fontenc}
\usepackage{ae,aecompl}
\usepackage{subfig}
\usepackage{booktabs}
\usepackage{longtable}
\usepackage{array}
\usepackage{pdflscape}
\usepackage{tabularx}
\usepackage{rotating}
\usepackage{multicol}
\usepackage{graphicx}
\restylefloat{figure}
\usepackage{subfig}
\usepackage{relsize}
\usepackage{graphicx}	
\usepackage{amsmath}
\usepackage{amssymb}	
\usepackage{enumitem}
\usepackage{stfloats}

\usepackage{colortbl}

\usepackage[customcolors]{hf-tikz}

\hfsetfillcolor{highlight!0}
\hfsetbordercolor{highlight!120}
\definecolor{highlight}{RGB}{163,17,71}
\definecolor{pastel}{RGB}{32, 91, 243}
\definecolor{spray}{RGB}{163,17,71}
\definecolor{telekom}{RGB}{219, 51, 210}
\definecolor{ultramarin}{RGB}{0,32,96}
\definecolor{lightgray}{gray}{0.75}
\definecolor{lightgray1}{gray}{0.85}
\definecolor{lightgray2}{gray}{0.8}
\definecolor{JonasColour}{RGB}{116, 4, 45}
\definecolor{JanColour}{RGB}{163,17,71}
\definecolor{DominikColour}{RGB}{0,100,0}

\usepackage{xcolor}
\definecolor{highlight}{RGB}{163,17,71}

\newcommand{\lambdabar}{{\mkern0.75mu\mathchar '26\mkern -9.75mu\lambda}}
\newcommand{\eqb}{\begin{equation}}
\newcommand{\eqe}{\end{equation}}
\newcommand{\dmb}{\begin{displaymath}}
\newcommand{\dme}{\end{displaymath}}

\newcommand{\eab}{\begin{eqnarray}}
\newcommand{\eae}{\end{eqnarray}}

\newcommand{\be}{\begin{equation}}
\newcommand{\ee}{\end{equation}}

\newcommand{\nn}{\newline}

\newcommand{\CMB}{\textnormal{\tiny \textsc{cmb}}}

\newcolumntype{C}{>{$\displaystyle} c <{$}}

\title{Axial Anomaly in Galaxies and the Dark Universe}

\author{ \href{https://orcid.org/0000-0001-7582-3456}{Janning Meinert\hspace{1mm}\includegraphics[scale=0.06]{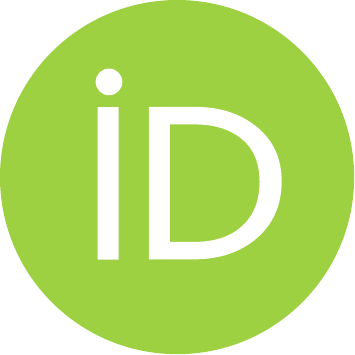}}\\
    Universit\"at Heidelberg,\\
    Institut f\"ur Theoretische Physik,\\
	Philosophenweg 16,\\
	D-69120 Heidelberg, Germany\\
	\texttt{j.meinert@thphys.uni-heidelberg.de} \\
	\And
	{Ralf Hofmann} \\
    Universit\"at Heidelberg,\\
    Institut f\"ur Theoretische Physik,\\
	Philosophenweg 16,\\
	D-69120 Heidelberg, Germany\\
	\texttt{r.hofmann@thphys.uni-heidelberg.de} \\
}

\begin{document}
\maketitle
\setcitestyle{square}

\begin{abstract}
Motivated by the SU(2)$_{\rm CMB}$ modification of the cosmological model $\Lambda$CDM, we consider isolated fuzzy-dark-matter lumps, made of ultralight axion particles whose masses arise due to distinct SU(2) Yang-Mills scales and the Planck mass $M_P$. In contrast to SU(2)$_{\rm CMB}$, these Yang-Mills theories are in confining phases (zero temperature) throughout most of the Universe's history and associate with the three lepton flavours of the Standard Model of particle physics. As the Universe expands, axionic fuzzy dark matter comprises a three-component fluid which undergoes certain depercolation transitions when dark energy (a global axion condensate) is converted into dark matter. We extract the lightest axion mass $m_{a,e}= 0.675\times 10^{-23}\,$eV from well motivated model fits to observed rotation curves in low-surface-brightness galaxies (SPARC catalogue). Since the virial mass of an isolated lump solely depends on $M_P$ and the associated Yang-Mills scale the properties of an e-lump predict those of $\mu$- and $\tau$-lumps. As a result, a typical e-lump virial mass $\sim 6.3\times 10^{10}\,M_\odot$ suggests that massive compact objects in galactic centers such as Sagittarius A$^*$ in the Milky Way are (merged) $\mu$- and $\tau$-lumps. In addition, $\tau$-lumps may constitute globular clusters. SU(2)$_{\rm CMB}$ is always thermalised, and its axion condensate never has depercolated. If the axial anomaly indeed would link leptons with dark matter and the CMB with dark energy then this would demystify the dark Universe through a firmly established feature of particle physics.
\end{abstract}

\keywords{galaxy rotation curves; low surface brightness; dark matter; dark energy; ultralight axion particles; cores; halos; mass-density; profiles; pure Yang-Mills theory}

%\begin{document}
\maketitle

\begin{multicols}{2}

%%%%%%%%%%%%%%%%%%%%%%%%%%%%%%%%%%%%%%%%%%%%%%%%%%\

%%%%%%%%%%%%%%%%% BODY OF PAPER %%%%%%%%%%%%%%%%%%\

\section{Introduction}

Dark matter was introduced as an explanation for the anomalous, kinematic behavior of luminous test matter 
in comparison with the gravity exerted by its luminous surroundings, e.g., virialised stars within a galaxy \cite{Rubin:1970zza} 
or a virialised galaxy within a cluster of galaxies \cite{Zwicky:1937zza}. That luminous matter can be segregated from dark 
matter is evidenced by the bullet cluster in observing hot intergalactic plasma (X-ray) in between 
localised dark-mass distributions (gravitational lensing) \cite{Tucker_1998,Clowe_2006}.\nn

The present Standard Model of Cosmology (SMC) $\Lambda$CDM posits a spatially flat Universe \cite{de_Bernardis_2000} with about 70\,\% dark energy, inducing late-time acceleration \cite{Riess_1998,Perlmutter:1998np}. This model requires a substantial contribution of about 26\,\% cold dark matter to the critical density and allows for a contribution of baryons of roughly 4\,\%.\nn

To determine all parameters of $\Lambda$CDM at a high accuracy, cosmological distance scales can be calibrated by high-redshift data (inverse distance ladder, global cosmology), coming from precision observations of the Cosmic Microwave Background (CMB) or from large-scale structure surveys probing Baryon Acoustic Oscillations (BAO). Alternatively, low-redshift data (direct distance ladder, local cosmology) can be used by appeal to standard or standardisable candles such as cepheids, TRGB stars, supernovae Ia, and supernovae II. Recently, a comparison between global and local cosmology has revealed tensions \cite{Verde_2019} in some of the cosmological parameter values (e.g., $H_0$ \cite{Aghanim:2018eyx,Shoes1,Shoes2,Wong:2016dpo} and $\sigma_8-\Omega_m$ \cite{DES1,DES2,Tr_ster_2020}, see also \cite{Hahn:2018dih} for the context of a high-redshift modification of $\Lambda$CDM).\newline

These interesting discrepancies motivate modifications of $\Lambda$CDM \cite{krishnan2021does}. A cosmological model aiming to resolve these tensions should target high-redshift radiation and the dark sector. In particular, models which are in principle falsifiable by terrestrial experiments and which pass such tests could lead to a demystification of the dark Universe. However, searches for weakly interacting, massive and stable particles (WIMPS) \cite{Kolb:1990vq}, whose potential existence is suggested by certain extensions of the Standard Model of Particle Physics (SMPP), so far have not produced any detection \cite{Akerib:2013tjd,Akerib:2018lyp}.\newline

An attractive possibility to explain the feebleness of a potential interaction between the dark sector of the SMC and SMPP matter in terms of the large hierarchy between particle-physics scales and the Planck mass is the theoretically \cite{Adler:1969er,Bell:1969ts,Fujikawa:1979ay} and experimentally \cite{Atherton:1985av} solidly anchored occurrence of an axial anomaly, which is induced by topological charge densities \cite{Peccei:1977ur} in the ground states of pure Yang-Mills theories \cite{bookHofmann}. The axial anomaly acts on top of a dynamical chiral symmetry breaking mediated by a force of hierarchically large mass scale compared to the scales of the Yang-Mills theories. To enable the axial anomaly throughout the Universe's entire history chiral fermions, which acquire mass through gravitational torsion and which can be integrated out in a Planck-scale de-Sitter background \cite{CandelasRaine}, need to be fundamentally charged under certain gauge groups. In such a scenario gravity itself - a strong force at the Planck scale - would induce the dynamical chiral symmetry breaking \cite{Frieman:1995pm,Gross:1973id,Giacosa:2008rw}. The anomaly then generates an axion mass $m_a$ \cite{Peccei:1977ur} for particles that a priori are chiral Nambu-Goldstone bosons. Working in natural units $c=\hbar=k_B=1$, one has

\begin{equation}
\label{maxxaxion}
m_a=\frac{\Lambda^2}{M_{P}}\,,
\end{equation}\nn
where $\Lambda$ denotes a Yang-Mills scale and $M_{P}=1.221\times10^{28}\,$eV the Planck mass \cite{Frieman:1995pm,Giacosa:2008rw}. 
The cold-dark-matter (CDM) paradigm is successful in explaining large-scale structure in the $\Lambda$CDM context but exhibits problems at small scales, e.g. galactic and lower \cite{Weinberg12249}: 
While N-body simulations within $\Lambda$CDM reveal matter-density profiles of the galactic DM halos that are characterised by a central cusp of the Navarro-Frenk-White (NFW) type \cite{Bullock:1999he}, $\rho_{\rm NFW}\propto r^{-1}$ \cite{Navarro:1996gj} ($r$ the radial distance to the center of the galaxy), observations suggest a core or soliton profile $\rho_{\rm sol}(r)$ subject to a constant central matter density $\rho_c=\rho_{\rm sol}(r=0)$, see e.g. \cite{Baldeschi:1983mq,Membrado:1989bqo,Sin1994,deBlok:2001rgg,KuziodeNaray:2007qi,Maleki:2020sqn,Pawlowski:2014aja}. A model of fuzzy dark matter (FDM) \cite{Baldeschi:1983mq,Sin1994,MatosGufman2000,Magana:2012ph,Suarez:2013iw,matos2016scalar,Marsh:2015xka,Hui:2016ltb,Schive:2014dra,Amorisco:2018dcn}, according to the ground-state solution of the Schr\"odinger-Poisson system embedded into cosmological simulations \cite{Schive:2014dra}, posits a condensate of free axion particles within the galactic core. For the radial range 
\begin{equation}
\label{relrerc}
r_{200}>r>r_e>3\,r_c
\end{equation}\nn
the associated central matter densities $\rho_{\rm sol}(r)$ gives way to a selfgravitating cloud of effective, nonrelativistic particles of mass $\sim \lambda_{\rm deB}^3\times \rho_{\rm NFW}(r)$. Here $r_{200}$ denotes the virial radius defined such that 
\begin{equation}
\label{NFWrange}
\rho_{\rm NFW}(r_{200})=200\,\frac{3\,M_P^2}{8\pi}H_0^2\,,
\end{equation}\nn
where $H_0$ is the Hubble constant, and $\lambda_{\rm deB}=\lambda_{\rm deB}(r)$ indicates 
the de-Broglie wavelength of an axionic particle for $r_e<r<r_{200}$ where the NFW model applies. Note that within the core region $r<r_e$ the correlation length in the condensate is given by the reduced Compton 
wave length $\lambdabar_C=1/m_a$. In what follows, we will refer to such a system -- condensate core plus NFW-tail -- as a {\sl lump}. In \cite{Bernal2017} FDM fits to the rotations 
curves of low-surface-brightness galaxies, which are plausibly assumed to be dominated by dark matter, have produced an axion mass of $m_a=0.554\times 10^{-23}\,$eV. Note also that the cosmological simulation of \cite{Schive:2014dra}  associates the axionic scalar field with dark-matter {\sl perturbations} only but not with the {\sl background} dark-matter density which is assumed to be conventional CDM. 

Another potential difficulty with $\Lambda$CDM, which FDM is capable of addressing, is the prediction of too many 
satellite galaxies around large hosts like the Milky Way or Andromeda \cite{Pawlowski:2012ft}, see, however, \cite{Nadler:2019hjw} for a cosmological simulation within CDM. A recent match of observed satellite abundances with 
cosmological simulations within the FDM context yields a stringent bound on the axionic particle mass $m_a$ \cite{Nadler:2019hjw}: $m_a>2.9\times 10^{-21}\,$eV. This bound is consistent with $m_a=2.5^{+3.6}_{-2.0}\times 10^{-21}\,$eV derived from an analysis of the Milky-Way rotation curve in \cite{Maleki:2020sqn}. 

There is yet another indication that $\Lambda$CDM may face a problem in delaying the formation of large galaxies of mass $M\sim 10^{12}\,M_{\odot}$ due to their hierarchical formation out of less massive ones. This seems to contradict the high-redshift observation of such galaxies \cite{Caputi:2015nra} and suggests that a component of active structure formation is at work. 

Assuming axions to be a classical ideal gas of non-relativistic particles the mass $m_a$ can be extracted from CMB simulations of the full Planck data subject to scalar adiabatic, isocurvature, and tensor-mode initial conditions \cite{Hlo_ek_2018} ($10^{-25}\,$eV$\le m_a\le 10^{-24}\,$eV with a 10\,\% contribution to DM and a 1\,\% contribution of isocurvature and tensor modes) and from a modelling of Lyman-$\alpha$ data \cite{Irsic:2017yje} with conservative assumptions on the thermal history of the intergalactic medium. For the XQ-100 and HIRES/MIKE quasar spectra samples one obtains respectively $m_a\ge 7.12\times 10^{-22}\,$eV and $m_a\ge 1.43\times 10^{-21}\,$eV. 

In our discussion of Sec.\,\ref{Sec5} we conclude that three axion species of hierarchically different masses could determine the dark-matter physics of our Universe. When comparing the results of axion-mass extractions with FDM based axion-mass constraints obtained in the literature it is important to observe that a {\sl single} axion species always is assumed. For example, this is true of the combined axion-mass bound $m_a>3.8\times 10^{-21}\,$eV, derived from modelling the Lyman-$\alpha$ flux power spectrum by hydrodynamical simulations \cite{Irsic:2017yje}, and it applies to the cosmological evolution of scalar-field based dark-matter perturbations yielding an axion mass of $m_a\sim 8\times 10^{-23}\,$eV in \cite{Schive:2014dra}. \newline

In the present article we are interested in pursuing the consequences of 
FDM for the physics of dark matter on super-galactic and sub-galactic scales within a cosmological model which deviates from $\Lambda$CDM in three essential points: 
(i) FDM is subject to three instead of one nonthermal axionic particle species, whose present cosmological mass densities are nearly equal, (ii) axion lumps (condensate core plus halo of fluctuating density granules) cosmologically originate from depercolation transitions at distinct redshifts $z_{p,i}$  out of homogeneous condensates  \cite{Hahn:2018dih}, and (iii) the usual, nearly scale invariant spectrum of adiabatic curvature fluctuations imprinted as an initial condition for cosmological cold-dark-matter evolution, presumably created by inflation, does not apply.\nn 

Point (i) derives from the match 
of axion species with the three lepton families of the Standard Model of 
particle physics. These leptons emerge in the confining phases of SU(2) Yang-Mills theories \cite{Hofmann:2017lmu}. According to Eq.\,(\ref{maxxaxion}) axion masses are then determined by the universal Peccei-Quinn 
scale $M_P$ and the distinct Yang-Mills scales $\Lambda_e$, $\Lambda_\mu$, and $\Lambda_\tau$.\nn  

Point (ii) is suggested by a cosmological model \cite{Hahn:2018dih} which is induced by the postulate that the CMB itself is described by an SU(2) gauge theory \cite{bookHofmann} and which fits the CMB power spectra TT, TE, and EE remarkably well except for low $l$. The according overshoot in TT at large angular scales may be due to the neglect of the nontrivial, SU(2)-induced photon dispersion at low frequencies.  
\nn   

Point (iii) relates to the fact that a condensate does not maintain density perturbations on cosmological scales and that $z_{p,e}\sim 53$. As a consequence, constraints on axion masses from cosmological simulations by confrontation with the observed small-scale structure should be repeated based on the model of \cite{Hahn:2018dih}. This, however, is beyond the scope of the present work.\nn 

To discuss point (ii) further, we refer to \cite{Hahn:2018dih}, where a dark sector was introduced as a deformation of $\Lambda$CDM. This modification models a 
sudden transition from dark energy to dark matter at a redshift $z_{\rm p}=53$. Such a transition is required phenomenologically to reconcile high-$z$ cosmology (well below the Planckian regime but prior to and including recombination), where the dark-matter density is reduced compared to $\Lambda$CDM, with well-tested low-$z$ cosmology. That a reduced dark-matter density is required at high $z$ is as a result of an SU(2)$_{\rm CMB}$-induced temperature-$z$ relation
\cite{Hofmann:2014lka}. Depercolation of a formely spatially homogeneous axion condensate, which introduces a change of the equation of state from $\rho=-P$ to $P=0$, is a result of the Hubble radius $r_H$ -- the spatial scale of causal connectedness in a Friedmann-Lemaitre-Robertson-Walker (FLRW) Universe -- exceeding by far the gravitational Bohr radius $r_B$ of an isolated, spherically symmetric system of selfgravitating axion particles. The value of the ratio $r_H/r_B$ at depercolation so far is subject to phenomenological extraction, but should intrinsically be computable in the future by analysis of the Schr\"odinger-Poisson system in a thus linearly perturbed background cosmology whose dark sector is governed by axion fields subject to their potentials.  

Roughly speaking, at depercolation from an equation of state $\rho=-P$ the quantum correlations in the axionic system become insufficient to maintain the homogeneity of the formerly homogeneously Bose-condensed state. The latter therefore decays or depercolates into selfgravitating islands of axionic matter whose central regions continue to be spatially confined Bose condensates but whose peripheries are virialised, quantum correlated particle clouds of an energy density that decays rapidly in the distance $r$ to the gravitational center to approach the cosmological dark-sector density. On cosmological scales, each of these islands (lumps) can be considered a massive (nonrelativistic) 
particle by itself such that the equation of state of the associated ensemble becomes $P=0$: The density of lumps then dilutes as $a^{-3}$ where $a$ denotes the cosmological scale factor. For the entire dark sector we have 
\begin{equation}
\label{edmdef}
\scalebox{0.9}{$\Omega_{\rm ds}(z)=\Omega_{\Lambda}+\Omega_{\rm pdm,0}(z+1)^3+\Omega_{\rm edm,0} 
\left\{\begin{array}{cc}
(z+1)^3,  & z<z_{p,e} \\
(z_{p,e}+1)^3, & z\geq z_{p,e} \\
\end{array}\right.\,.
$}
\end{equation}\nn
Fits of this model to the TT, TE, and EE CMB power spectra reveal that $\Omega_{\rm edm,0} \sim \frac{1}{2} \Omega_{\rm pdm,0}$. Here $\Omega_{\rm pdm,0}$ denotes a primordial contribution to the present dark-matter density parameter $\Omega_{\rm dm,0}=\Omega_{\rm edm,0}+\Omega_{\rm pdm,0}$ while 
$\Omega_{\rm edm,0}$ refers to the emergence of dark matter due to the depercolation of a formerly homogeneous Bose-Einstein condensate into isolated lumps once their typical Bohr radius is well covered by the horizon radius $r_{H}$. One may question that depercolation occurs suddenly at $z_{p,e}$, the only justification so far being the economy of the model. 
If a first-principle simulation of the Schr\"odinger-Poisson system plus background cosmology reveals  
that the transition from dark energy to dark matter during depercolation involves a finite $z$-range then this has to be included in the model of Eq.\,(\ref{edmdef}).   

After depercolation has occurred, a small dark-energy 
residual $\Omega_{\Lambda}$ persists to become the dominant cosmological constant today. As we will argue in Sec.\,\ref{Sec5}, the primordial dark-matter density $\Omega_{\rm pdm,0}$ could originate from the stepwise depercolation of former dark energy in the form of super-horizon sized $\mu$- and $\tau$-lumps. Therefore, dark energy dominates the dark sector at sufficiently high $z$. However, due to radiation dominance dark energy then was a marginal contribution to the expansion rate. The model of \cite{Hahn:2018dih} was shown to fit the CMB anisotropies with a low baryon density, the local value for the redshift of re-ionisation \cite{Becker_2001}, and the local value of $H_0$ from supernovae Ia distance-redshift extractions \cite{Shoes1,Shoes2}.\nn

The purpose of the present work is to propose a scenario which accommodates $\Omega_{\rm edm,0}$, $\Omega_{\rm pdm,0}$, and $\Omega_{\Lambda}$. At the same time, we aim at explaining the parameters $\Omega_{\rm edm,0}$ and $\Omega_{\rm pdm,0}$ in terms of axial anomalies subject to a Planck-mass Peccei-Quinn scale and three SU(2) Yang-Mills theories associated with the three lepton families. In addition, an explanation of parameter $\Omega_{\Lambda}$ is proposed which invokes the SU(2) Yang-Mills theory underlying the CMB. Hence, the explicit gauge-theory content of our model is: SU(2)$_e\times\,$SU(2)$_{\mu}\times\,$SU(2)$_{\tau}\times\,$SU(2)$_{\rm CMB}\,$.\nn

We start with the observation in \cite{Sin1994} that ultralight bosons necessarily need to occur in the form of selfgravitating condensates in the cores of galaxies. Because these cores were separated in the course of nonthermal depercolation halos of axion particles, correlated due to gravitational virialisation on the scale of their de Broglie wavelength, were formed around the condensates. Such a halo reaches out to a radius, say, of $r_{200}$ where its mass density starts to fall below 200 times the critical cosmological energy density of the spatially flat FLRW Universe. A key concept in describing such a system - a lump - is the gravitational Bohr radius $r_B$ defined as 
\begin{equation}
\label{Bohr}
r_B\equiv \frac{M_{\rm P}^2}{M m_a^{2}}\,, 
\end{equation}\nn
where $M$ is the mass of the lump which should coincide with the viral mass, say $M_{200}$. We use two FDM models of the galactic mass density $\rho(r)$ to describe low-surface-brightness galaxies and to extract the axion mass $m_a$: The Soliton-NFW model, see \cite{matos2016scalar} and references therein, and the Burkert model \cite{Burkert_1995, Salucci:2000ps}.

Rather model independently, we extract a typical value of $m_{a,e}\sim 0.7\times 10^{-23}$\,eV which confirms the value obtained in \cite{Bernal2017}. With Eq.\,(\ref{maxxaxion}) this value of $m_{a,e}$ implies a Yang-Mills scale of $\Lambda_e \sim 287$\,eV. This is smaller than $\Lambda_e=511\,{\rm keV}/118.6=4.31$\,keV found in \cite{Hofmann:2017lmu} where a link to an SU(2) Yang-Mills theory governing the first lepton family is made: SU(2)$_e$. Note that the larger value of $\Lambda_e$ was extracted in the {\sl deconfining} phase \cite{Hofmann:2017lmu} while the smaller value, obtained from the axion mass $m_{a,e}$, relates to the confining phase. The suppression of Yang-Mills scale is plausible because topological charges, which invoke the axial anomaly, are less resolved in the confining as compared to the deconfining phase. The gravitational Bohr radius associated with a typical e-lump mass of $M_e\sim 6.3\times 10^{10}\,M_{\odot}$ turns out to be $r_{B,e}\sim 0.26\,$kpc.\nn  

Having fixed the scales of SU(2)$_{\rm CMB}$, SU(2)$_{\rm e}$ and linked their lumps to dark energy 
and the dark-matter halos of low-surface-brightness galaxies, respectively, we associate the lumps of 
SU(2)$_{\rm \mu}$ and SU(2)$_{\rm \tau}$ with $\Omega_{\rm pdm,0}$ of the dark-sector cosmological model in Eq.\,(\ref{edmdef}). Within a galaxy, each individual $\mu$- and $\tau$-lump provides a mass fraction of $(m_e/m_\mu)^2\sim 2.3\times 10^{-5}$ and $(m_e/m_\tau)^2\sim 8.3\times 10^{-8}$, respectively, 
of the mass $M_e$ of an e-lump, see Eq.\,(\ref{lumpmassratios}).\nn 

This paper is organised as follows. In Sec.\,\ref{Sec2} we discuss features of lumps in terms of a universal ratio between reduced Compton wavelength and gravitational Bohr radius. As a result, a typical lump mass can be expressed solely in terms of Yang-Mills scale and Planck mass. The rotation curves of galaxies with low surface brightness (SPARC library) are analysed in Sec.\,\ref{Sec3} using two models with spherically symmetric mass densities: the Soliton-Navarro-Frenk-White (SNFW) and the Burkert model. Assuming that only one Planck-scale axion species dominates the dark halo of a low-surface-brightness galaxy in terms of an isolated, unmerged e-lump, we extract the typical axion mass $m_{a,e}$ in Sec.\,\ref{SNFW}. In Sec.\,\ref{BM} we demonstrate the consistency of axion-mass extraction between the two models: The gravitational Bohr radius, determined in SNFW, together with the lump mass, obtained from the Burkert-model-fit, predicts an axion mass which is compatible with the axion mass extracted from the soliton-core density of the SNFW model. The typical value of the axion mass suggests an association with SU(2) Yang-Mills dynamics responsible for the emergence of the first lepton family. In Sec.\,\ref{Sec4} this information is used to discuss the cosmological origin and role of lumps played in the dark Universe in association with the two other lepton families and the SU(2) gauge theory propounded to describe the CMB \cite{Hofmann:2014lka,Hahn:2018dih}. As a result, on subgalactic scales the $\mu$-lumps could explain the presence of massive compact objects in galactic centers such as Sagittarius A$^*$ in the Milky Way \cite{Gillessen:2008qv,Ghez2008} while $\tau$-lumps may relate to globular clusters \cite{Kalberla:2007sr}. On super-galactic scales and for $z<z_{p,e}$, however, lumps from all axion species act like CDM. On the other hand, the CMB-lump's extent always exceeds the Hubble radius by many orders of magnitude and therefore should associate with dark energy. Finally, in Sec.\,\ref{Sec5} we discuss in more detail how certain dark structures of the Milky Way may have originated in terms of $\mu$- and $\tau$-lumps. We also provide a summary and an outlook on future work. We work in natural units $\hbar=c=k_B=1$.

\medskip

\section{Gravitational Bohr radius and reduced Compton wave length of a Planck-scale axion}\label{Sec2}

We start by conveying some features of basic axion lumps, cosmologically originated by depercolation transitions, that we wish to study. Let 
\begin{equation}
\label{Compton}
\lambdabar_{C,i}\equiv \frac{1}{m_{a,i}}
\end{equation}\nn
denote the reduced Compton wavelength and  
\begin{equation}
\label{intpartD}
d_{a,i}\equiv \left(\frac{m_{a,i}}{\bar{\rho}_{i}}\right)^{1/3} 
\end{equation}\nn
the mean distance between axion particles within the spherically symmetric core of the lump of mean dark-matter mass density $\bar{\rho}_{i}$. One has 
\begin{equation}
\label{densmean}
\bar{\rho}_{i}\sim\frac{M_{200,i}}{\frac{4\pi}{3}r_{B,i}^3}\,. 
\end{equation}\nn
The energy densities $\rho_i$ of each of the three dark-energy like homogeneous condensates of axionic particles prior to lump depercolation 
are assumed to arise due to Planckian physics \cite{Giacosa:2008rw}. Therefore, each $\bar{\rho}_i$ may only depend on $M_P$ and $m_{a,i}$ ($i=e,\mu,\tau$). Finite-extent, isolated, unmerged lumps self-consistently are characterised by a fixed ratio between the reduced Compton wavelength $\lambdabar_{C,i}$ - the correlation length in the condensate of free axion particles at zero temperature - and the Bohr radius $r_{B,i}$.

Let us explain this. Causal lump segregation due to cosmological expansion (depercolation), which sets in when the Hubble radius $r_H$ becomes sufficiently larger than $r_B$, is adiabatically slow and generates a sharply peaked distribution of lump masses (and Bohr radii) in producing typically sized condensate cores. These cores are surrounded by halos of axion particles that represent regions of the dissolved condensate and nonthermally are released by the mutual pull of cores during depercolation. In principle, we can state that for an isolated, unmerged lump
\vspace{4mm}
\begin{equation}
\label{CvB}
\frac{r_{B,i}}{\lambdabar_C}=\kappa(\delta_i)\,,
\end{equation}
where $\kappa$ is a smooth dimensionless function of its dimensionless argument $\delta_i\equiv m_{a,i}/M_P$ with the property that $\lim_{\delta_i\to 0}\kappa_i(\delta_i)<\infty$. This is because the typical mass $M_i\sim M_{200,i}$ of an isolated, unmerged lump, which enters $r_{B,i}$ via Eq.\,(\ref{Bohr}), is, due to adiabatically slow depercolation, by itself only a function of the two mass scales $m_{a,i}$ and $M_P$ mediating the interplay between quantum and gravitational correlations that give rise to the formation of the lump. Since $\delta_i$ is much smaller than unity, we can treat the right-hand side of Eq.\,(\ref{CvB}) as a {\sl universal} constant. In practice, we will in Sec.\,\ref{Sec3} derive the values of $r_{B,e}$ and $m_{a,e}$ by matching dark-matter halos of low surface-brightness galaxies with well motivated models of a lump's mass density. As a result, we state a value of $\kappa\sim 314$ in Eq.\,(\ref{kappavalue}) of Sec.\,\ref{Sec4}. 

Eq.\,(\ref{CvB}) together with Eqs.\,(\ref{Compton}), (\ref{Bohr}), and (\ref{maxxaxion}) imply for the mass $M_i$ of the isolated, unmerged lump 
 \begin{equation}
\label{MLambda}
M_i=\frac{1}{\kappa}\frac{M_P^3}{\Lambda_i^2}\,.
\end{equation}\nn
Eq.\,(\ref{MLambda}) is important because it predicts that the ratios of lump masses solely are determined by the squares of the ratios of the respective Yang-Mills scales or, what is the same \cite{Hofmann:2017lmu}, by the ratios of charged lepton masses $m_e$, $m_\mu$, and $m_\tau$. One has 
\eab
\label{lumpmassratios}
\frac{M_\tau}{M_\mu}&=&\left(\frac{m_\tau}{m_\mu}\right)^2\sim 283\,,\nonumber\\
\frac{M_\mu}{M_e}&=&\left(\frac{m_e}{m_\mu}\right)^2\sim 2.3\times 10^{-5}\,,\\ 
\frac{M_\tau}{M_e}&=&\left(\frac{m_e}{m_\tau}\right)^2\sim 8.3\times 10^{-8}\,.\nonumber
\eae
Moreover, Eqs.\,(\ref{maxxaxion}), (\ref{intpartD}), (\ref{Compton}), (\ref{densmean}), and (\ref{MLambda}) 
fix the ratio $\xi_i\equiv\frac{d_{a,i}}{\lambdabar_{C,i}}$ as 
 \begin{equation}
\label{xidef}
\xi_i=\left(\frac{4\pi}{3}\right)^{1/3}\left(\kappa\frac{\Lambda_i}{M_P}\right)^{4/3}\,.
\end{equation}\nn
Since $\Lambda_i\ll M_P$ we have $\xi_i\ll 1$, and therefore a large number of axion 
particles are covered by one reduced Compton wave length. This assures that the assumption of a condensate core is selfconsistent. A thermodynamical argument for the necessity of axion condensates throughout the Universe's expansion history is given in Sec.\,\ref{Sec4}. In \cite{Sin1994} the non-local and non-linear 
(integro-differential) Schr\"odinger-equation, obtained from a linear Schr\"odinger equation and a Poisson equation for the gravitational potential, see e.g. \cite{Bar_2018}, governing the lump, was analysed. An excitation of such a lump in terms of its wave-function $\psi_i$ containing radial zeros 
was envisaged in \cite{Sin1994,Bernal2017}. Here instead, we assume the isolated, unmerged lump to be in its ground 
state, parameterised by a phenomenological mass density 
$\rho_i(r)\propto |\psi_i|^2(r)>0$ which represents the lump well \cite{Schive:2014dra}.

Finally, Eq.\,(\ref{Bohr}) together with Eqs.\,(\ref{maxxaxion}) and (\ref{MLambda}) yield for the 
gravitational Bohr radius
 \begin{equation}
\label{Bohrradiusfund}
r_{B,i}=\kappa\frac{M_P}{\Lambda_i^2}\,.
\end{equation}

\section{Analysis of rotation curves}\label{Sec3}

In this section, we extract the axion mass $m_{a,e}$ from observed RCs of low-surface-brightness galaxies which fix the lump mass $M_e$ and a characterising length scale -- the gravitational Bohr radius $r_{B,e}$. This, in turn, determines the (primary, see Sec.\,\ref{Sec4}) Yang-Mills scale $\Lambda_e$ associated with the lump. We analyse RCs from the 
SPARC library \cite{Lelli:2016zqa}. 

\subsection{Fuzzy Dark Matter: SNFW vs. Burkert model}

To investigate, for a given galaxy and RC, the underlying spherically symmetric 
mass density $\rho(r)$ it is useful to introduce the orbit-enclosed mass
 \begin{equation}
M(r) = 4\pi\int_{0}^{r}dr'\, r'^2\rho_{{\rm }}(r')\,.
\label{eq:Mtot}
\end{equation}\nn
Assuming virialisation, spherical symmetry, and Newtonian gravity the orbital 
velocity $V(r)$ of a test mass (a star) is given as   
\begin{align}
V(r) = \sqrt{\frac{G M(r)}{r}}\,,
\label{eq:Vcirc}
\end{align}\nn
where $M(r)$ is defined in Eq.\,(\ref{eq:Mtot}), and $G\equiv M_P^{-2}$ denotes Newton's constant. The lump mass 
$M$ is defined to be $M_{200}\equiv M(r_{200})$ where $r_{200}$ is given by Eq.\,(\ref{NFWrange}). For an extraction of $m_{a,e}$ and therefore the associated Yang-Mills scale governing the mass of a lump according to Eq.\,(\ref{MLambda}), we use the Soliton-Navarro-Frenk-White (SNFW) 
and the Burkert model. The mass-density profile of the NFW-part of the SNFW-model is  
given as \cite{Navarro:1996gj}
%Robles:2012uy
\begin{align}
\rho_{{\rm NFW}}(r) = \frac{\rho_s^{\rm NFW}}{\frac{r}{r_s}(1+\frac{r}{r_s})^2}\,,
\label{eq:RhoNFW}
\end{align}\nn
where $\rho_s^{\rm NFW}$ associates with the central mass density, and $r_s$ is a scale radius which represents the onset of the asymptotic cubic decay in distance $r$ to the galactic center. Note that profile $\rho_{\rm NFW}$ exhibits an infinite cusp as $r\to 0$ and that the orbit-enclosed mass $M(r)$ diverges logarithmically with the cutoff radius $r$ for the integral in Eq.\,(\ref{eq:Mtot}). In order to avoid the cuspy behavior for $r\to 0$, an axionic Bose-Einstein condensate (soliton density profile) is assumed to describe the soliton region $r\le r_e$. From the ground-state solution of the Schr\"odinger-Poisson system for a 
single axion species one obtains a good analytic description of the soliton density profile as \cite{Bernal2017}
\begin{align}
\rho_{{\rm sol}}(r) = \frac{\rho_c}{(1+0.091(r/r_c)^2)^8}\,,
\label{eq:SolitonNFW}
\end{align}
where $\rho_c$ is the core density \cite{Schive:2014dra}. On the whole, the fuzzy dark matter profile can than be approximated as 
\begin{align}
\rho_{{\rm FDM}}(r) = \Theta(r_\epsilon-r)\rho_{{\rm sol}}+ \Theta(r-r_\epsilon)\rho_{{\rm NFW}}\,.
\label{eq:SolitonNFWwhole}
\end{align}
For the Burkert model one assumes a mass-density profile of the form \cite{Burkert_1995, Salucci:2000ps}
%Robles:2012uy
 \begin{equation}
\label{eq:RhoPI}
\rho_{{\rm Bu}}(r) = \frac{\rho_0\, r_0^3}{(r+r_0)(r^2+r_0^2)}\,
\end{equation}
where $\rho_0$ refers to the central mass density and $r_0$ is a scale 
radius.

%\newpage

%########################## SNFW ###############################\

\subsection{Analysis of RCs in the SNFW model}\label{SNFW}

\begin{figure}[H]\centering
\includegraphics[width=7.5cm]{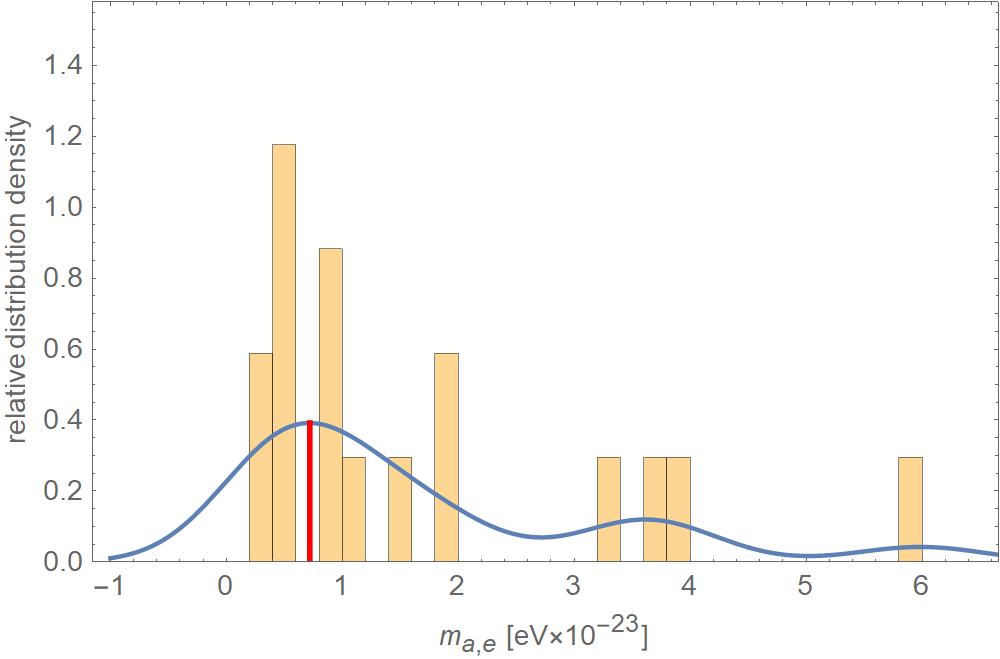}
\caption{Frequency distribution of axion mass $m_{a,e}$ as extracted from the SNFW model for 17 best fitting galaxies. The maximum of the smooth-kernel-distribution (solid, blue line) is at $m_{a,e}=(0.72\pm 0.5) \times10^{-23}$ eV  (red, vertical line).}
\label{fig:AxionMassesSNFW}
\end{figure}

Using Eqs.\, (\ref{eq:Mtot}), (\ref{eq:Vcirc}), and (\ref{eq:SolitonNFWwhole}), 
we obtain the orbital velocity $V_{\rm SNFW}$ of the 
SNFW model \cite[Eq. (17)]{Robles:2012uy} which is fitted to observed RCs. This determines the 
parameters $r_\epsilon$, $r_s$, and $\rho_c$. The density $\rho_s$ relates to these fit parameters by demanding continuity of the SNFW mass density at $r_\epsilon$ \cite{Bernal2017}. As a result, one has  
 \begin{equation}
\label{rhocrhos}
\rho_s(\rho_c,r_c,r_\epsilon,r_s)=\rho_c \frac{(r_\epsilon/r_s)(1+r_\epsilon/r_s)^2}{(1+0.091(r_\epsilon/r_c)^2)^8}\,.
\end{equation}\nn
Examples of good fits with $\chi^2/\text{d.o.f.}<1$ are shown in Table\,1, see Table\,2 and Table\,3 for the corresponding fit parameters. The derived quantity $m_{a,e}$ is extracted from the following equation \cite{Schive:2014dra}
 \begin{equation}
\label{mafromrhoc}
\rho_c \equiv 1.9\times 10^9 (m_{a,e}/10^{-23} {\rm eV})^{-2} (r_c/{\rm kpc})^{-4} M_{\odot}{\rm kpc}^{-3}
\end{equation}\nn
The other derived quantities $r_{200}$ and $M_{200}$ are obtained by employing Eqs.\,(\ref{NFWrange}) and (\ref{eq:Mtot}) with $M(r=r_{200})\equiv M_{200}$, respectively. In Fig.\,\ref{fig:AxionMassesSNFW} a frequency distribution of $m_{a,e}$ is shown, based on a sample 
of 17 best fitting galaxies, see Table\,1 for the fits to the RCs. The maximum of the smooth-kernel-distribution (solid blue line in Fig.\,\ref{fig:AxionMassesSNFW}) is at 
%$m_{a,e}=(0.72\pm 0.5)\times 10^{-23}\,{\rm eV}\,$.
 \begin{equation}
m_{a,e}=(0.72\pm 0.5)\times 10^{-23}\,{\rm eV}\,.
\end{equation}\nn

%###################### SNFW Fits ##########################\

\begin{figure*}
\textbf{Table 1.} Best fits of SNFW to RCs of 17 SPARC galaxies. The arrows indicate the Bohr radius of the e-lump, $r_{B,e}$ (red), the core radius of the soliton $r_{c}$ (orange), the transition radius from the soliton model to the NFW model $r_{e}$ (yellow), and the scale radius of the NFW model $r_{s}$ (green).

\vspace{0.3cm}
\begin{minipage}{1\textwidth}
\centering
\hspace{-5mm}\includegraphics[width=17cm]{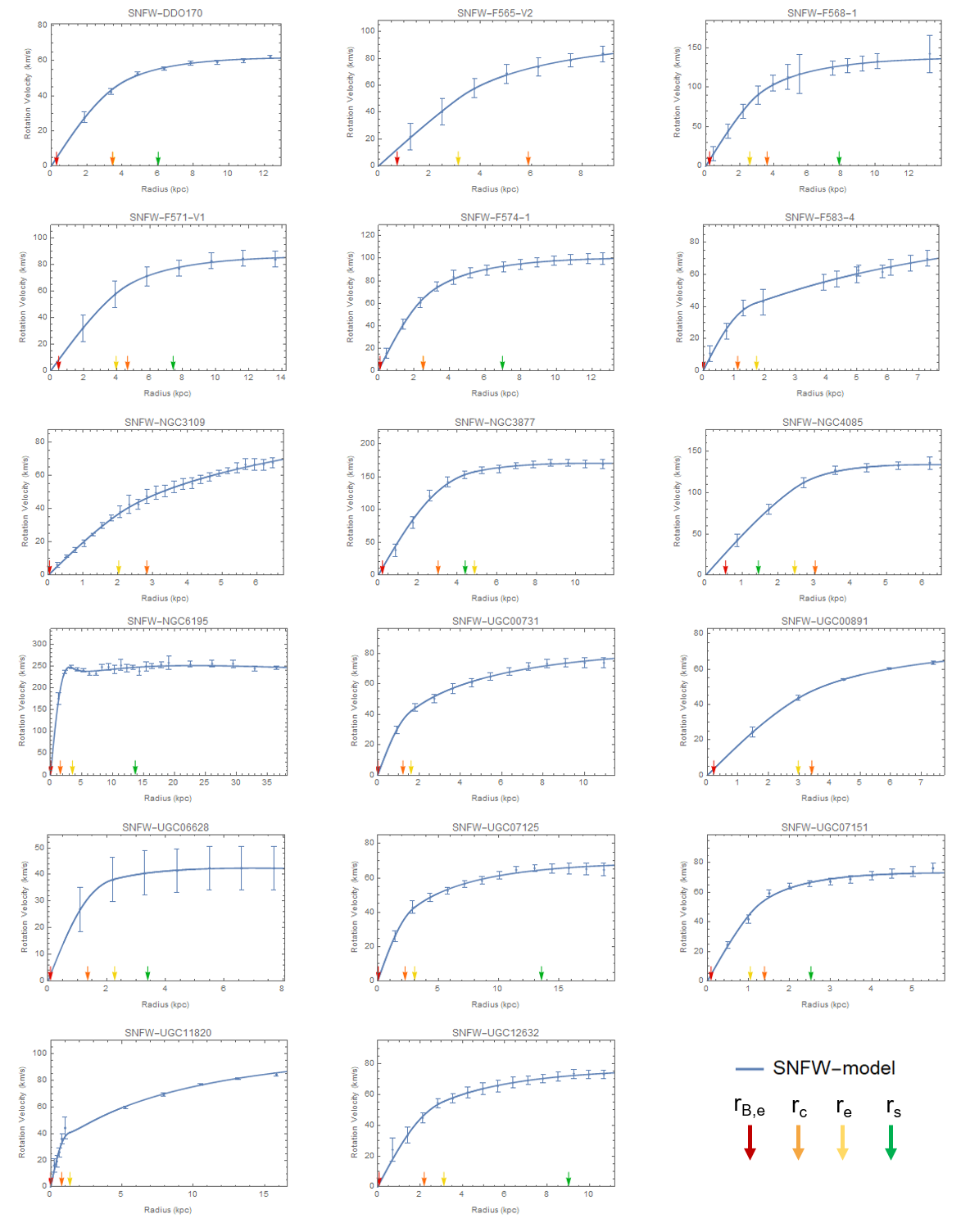}
\end{minipage}
%\label{GalaxyMasses}
\end{figure*}

%\newpage

\end{multicols}{}

\nointerlineskip
    \begin{table}[H]
    %\tablesize{\scriptsize}
    %\widetable
\caption*{{\bf Table 2}. Fits of RCs to SNFW model: Galaxy name, Hubble Type, $\chi^2/\text{d.o.f.}$, luminosity, axion mass $m_a$, $r_{200}$, virial mass $M_{200}$,  central density $\rho_c$, scale radius $r_s$, transition radius $r_e$, core radius $r_c$, and $r_e/r_c$. The fitting constraints are heuristic and motivated by the results of \cite{Bernal2017}, $r_{200} < 200$\,kpc, $r_e, r_c<$ 6\,kpc, and $r_e/r_c$ > 0.1.}\label{table2fig:GalaxyMassesSNFW}

%######################## first part of former table 2 ####

%{\small
\[
\begin{array}{|l|c|c|c|c|c|c|}
\hline
 \text{Galaxy} & 
\begin{array}{c}
 \text{Hub.} \\
 \text{Type} \\
\end{array}
 & 
\begin{array}{c}
 \text{Lum.} \\
 \left[L_{\odot}\right/\text{pc}^2] \\
\end{array}
 & 
\begin{array}{c}
 \text{$\chi $2 } \\
 \text{/d.o.f.} \\
\end{array}
 & 
\begin{array}{c}
 m_{a,e} \\
 \left.\text{[eV$\times $}10^{-23}\right] \\
\end{array}
 & 
\begin{array}{c}
 r_{200} \\
 \text{[kpc]} \\
\end{array}
 & 
\begin{array}{c}
 M_{200}  \\
 \left[M_{\odot}\times 10^{10}\right] \\
\end{array}
 \\\hline
%--------------------------- SNFW ---------------------
 \text{DDO170} & Im & 73.93 & 0.73 & 1.00\pm0.82 & 36.90\pm8.16 & 2.81\pm1.95 \\\hline
 \text{F565-V2} & \text{Im} & 40.26 & 0.02 & 0.31\pm0.47 & 60.71\pm11.52 & 11.65\pm5.05 \\\hline
 \text{F568-1} & \text{Sc} & 57.13 & 0.02 & 0.41\pm0.24 & 70.06\pm7.82 & 22.59\pm6.67 \\\hline
 \text{F571-V1} & \text{Sd} & 64.39 & 0.03 & 0.50\pm0.30 & 49.71\pm6.70 & 7.12\pm2.83 \\\hline
 \text{F574-1} & \text{Sd}& 128.48 & 0.02 & 0.93\pm0.15 & 53.96\pm3.09 & 9.79\pm1.44 \\\hline
 \text{F583-4} & \text{Sc}& 83.34 & 0.13 & 3.87\pm1.37 & 75.08\pm32.44 & 17.88\pm13.83 \\\hline
 \text{NGC3109} & \text{Sm}& 140.87 & 0.18 & 1.14\pm0.39 & 77.09\pm18.07 & 19.85\pm5.45 \\\hline
 \text{NGC3877} & \text{Sc}& 3410.59 & 0.17 & 0.39\pm0.04 & 69.37\pm12.49 & 26.91\pm6.58 \\\hline
 \text{NGC4085} & \text{Sc}& 5021.46 & 0.07 & 0.41\pm0.18 & 46.07\pm7.09 & 9.26\pm2.78 \\\hline
 \text{NGC6195} & \text{Sb}& 174.11 & 0.47 & 0.48\pm0.03 & 121.72\pm9.61 & 122.40\pm24.56 \\\hline
 \text{UGC00731} & \text{Im}& 82.57 & 0.19 & 3.39\pm0.87 & 53.27\pm8.51 & 7.62\pm2.90 \\\hline
 \text{UGC00891} & \text{Sm}& 113.98 & 0.01 & 0.93\pm0.10 & 44.92\pm1.42 & 4.78\pm0.36 \\\hline
 \text{UGC06628} & \text{Sm}& 103.00 & 0.00 & 3.61\pm0.10 & 23.35\pm0.87 & 0.77\pm0.05 \\\hline
 \text{UGC07125} & \text{Sm}& 103.00 & 0.25 & 1.81\pm0.65 & 47.55\pm10.38 & 4.97\pm2.68 \\\hline
 \text{UGC07151} & \text{Scd} & 965.67 & 0.72 & 1.94\pm2.32 & 32.32\pm5.65 & 2.53\pm1.58 \\\hline
 \text{UGC11820} & \text{Sm} & 34.11 & 0.82 & 5.99\pm4.75 & 75.83\pm49.15 & 18.10\pm34.72 \\\hline
 \text{UGC12632} & \text{Sm} & 66.81 & 0.09 & 1.47\pm0.23 & 46.91\pm6.10 & 5.56\pm1.48 \\\hline
\end{array}
\]
%}  %for the small environment

\end{table}

\vspace{0.5cm}

%############### second part of former table 2 (now table 3) ####

\nointerlineskip
    \begin{table}[H]
\caption*{{\bf Table 3}. Fits of RCs to SNFW model: Galaxy name, central density $\rho_c$, soliton density $\rho_s$, scale radius $r_s$, transition radius $r_e$, core radius $r_c$, and $r_e/r_c$. The fitting constraints are heuristic and motivated by the results of \cite{Bernal2017}, $r_{200} < 200$\,kpc, $r_e, r_c<$ 6\,kpc, and $r_e/r_c$ > 0.1.}\label{table2-B-fig:GalaxyMassesSNFW}\vspace{12pt}

%{\small
\[
\begin{array}{|l|c|c|c|c|c|c|}\hline
 \text{Galaxy} & 
\begin{array}{c}
 \rho _C\times 10^7 \\
 \left[M_{\odot}\right/\text{kpc}^3] \\
\end{array}
 & 
\begin{array}{c}
 \rho _s\times 10^7 \\
 \left[M_{\odot}\right/\text{kpc}^3] \\
\end{array}
 & 
\begin{array}{c}
 r_s \\
 \text{[kpc]} \\
\end{array}
 & 
\begin{array}{c}
 r_e \\
 \text{[kpc]} \\
\end{array}
 & 
\begin{array}{c}
 r_c \\
 \text{[kpc]} \\
\end{array}
 & r_e/r_c \\\hline
 %--------------------------- SNFW ---------------------
 \text{DDO170} & 1.34\pm0.38 & 0.95\pm0.78 & 6.02\pm1.51 & 3.48\pm2.99 & 3.45\pm1.39 & 1.01 \\\hline
 \text{F565-V2} & 1.60\pm0.21 & 0.51\pm0.33 & 12.46\pm4.33 & 3.14\pm0.80 & 5.89\pm4.38 & 0.53 \\\hline
 \text{F568-1} & 6.61\pm0.56 & 2.67\pm1.04 & 7.87\pm1.36 & 2.59\pm0.60 & 3.62\pm1.06 & 0.72 \\\hline
 \text{F571-V1} & 1.61\pm0.24 & 1.22\pm0.68 & 7.42\pm1.71 & 3.96\pm1.21 & 4.65\pm1.40 & 0.85 \\\hline
 \text{F574-1} & 5.41\pm0.26 & 1.81\pm0.32 & 6.98\pm0.46 & 2.52\pm1.04 & 2.52\pm0.21 & 1.00 \\\hline
 \text{F583-4} & 8.05\pm1.34 & 0.12\pm0.13 & 27.10\pm19.49 & 1.73\pm0.41 & 1.12\pm0.19 & 1.55 \\\hline
 \text{NGC3109} & 2.28\pm0.09 & 0.14\pm0.08 & 25.88\pm11.92 & 2.03\pm0.56 & 2.83\pm0.48 & 0.71 \\\hline
 \text{NGC3877} & 15.00\pm0.91 & 13.56\pm15.70 & 4.40\pm2.43 & 4.87\pm0.91 & 3.03\pm0.15 & 1.61 \\\hline
 \text{NGC4085} & 13.49\pm1.04 & 103.33\pm134.83 & 1.46\pm0.82 & 2.46\pm0.41 & 3.02\pm0.65 & 0.81 \\\hline
 \text{NGC6195} & 128.10\pm7.47 & 2.67\pm0.60 & 13.68\pm0.96 & 3.55\pm0.12 & 1.59\pm0.05 & 2.23 \\\hline
 \text{UGC00731} & 7.41\pm0.87 & 0.39\pm0.16 & 12.17\pm1.85 & 1.62\pm0.34 & 1.22\pm0.15 & 1.32 \\\hline
 \text{UGC00891} & 1.63\pm0.05 & 0.57\pm0.06 & 8.85\pm0.39 & 2.97\pm0.19 & 3.42\pm0.19 & 0.87 \\\hline
 \text{UGC06628} & 4.34\pm0.08 & 1.30\pm0.21 & 3.40\pm0.28 & 2.27\pm0.07 & 1.35\pm0.02 & 1.68 \\\hline
 \text{UGC07125} & 2.20\pm0.39 & 0.22\pm0.12 & 13.55\pm2.20 & 3.06\pm0.94 & 2.27\pm0.39 & 1.35 \\\hline
 \text{UGC07151} & 13.48\pm2.86 & 7.49\pm5.62 & 2.52\pm0.66 & 1.05\pm0.46 & 1.39\pm0.83 & 0.75 \\\hline
 \text{UGC11820} & 15.04\pm4.43 & 0.11\pm0.20 & 28.64\pm6.01 & 1.35\pm0.66 & 0.77\pm0.30 & 1.76 \\\hline
 \text{UGC12632} & 3.77\pm0.30 & 0.61\pm0.24 & 9.01\pm1.82 & 3.12\pm0.49 & 2.20\pm0.16 & 1.42 \\\hline
\end{array}
\]
%}  %for the small environment

\end{table}
\clearpage

\begin{multicols}{2}

%--------------------------------------------

\begin{figure}[H]\centering
\includegraphics[width=7.5cm]{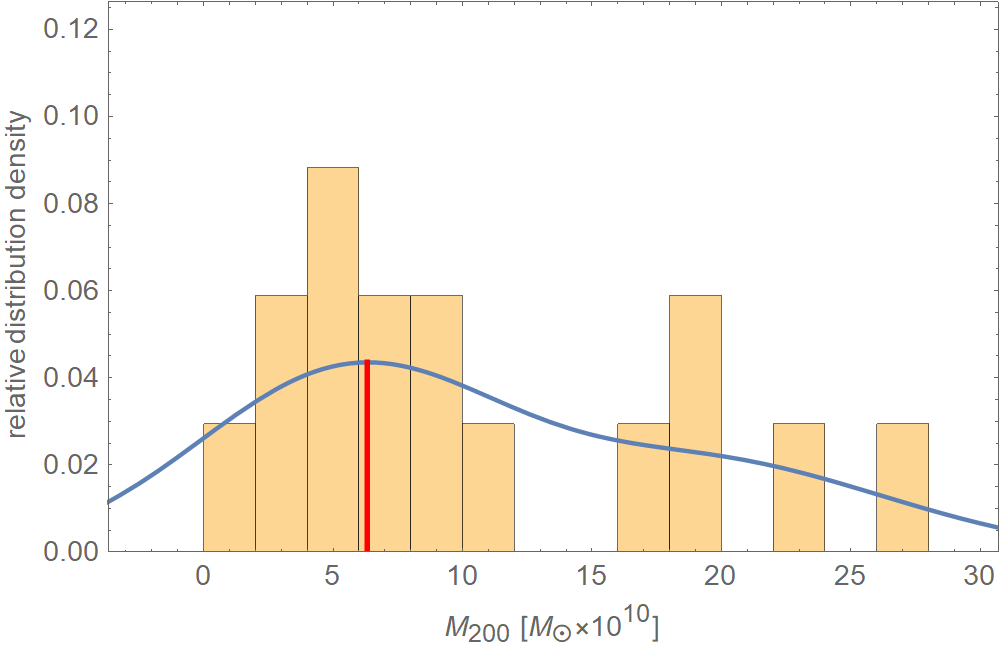}
\caption{
Frequency distribution of the virial mass $M_{200}$ in units of solar masses $M_{\odot}$ from the 17 best-fitting galaxies in the SNFW model. The maximum %of the smooth-kernel-distribution (solid line) 
of the distribution is at $M_{200}=(6.3\pm 4)\times 10^{10}\,M_{\odot}$.
}
\label{fig:HistoMassesSNFW}
\end{figure}

\begin{figure}[H]
\centering
\includegraphics[width=7.5cm]{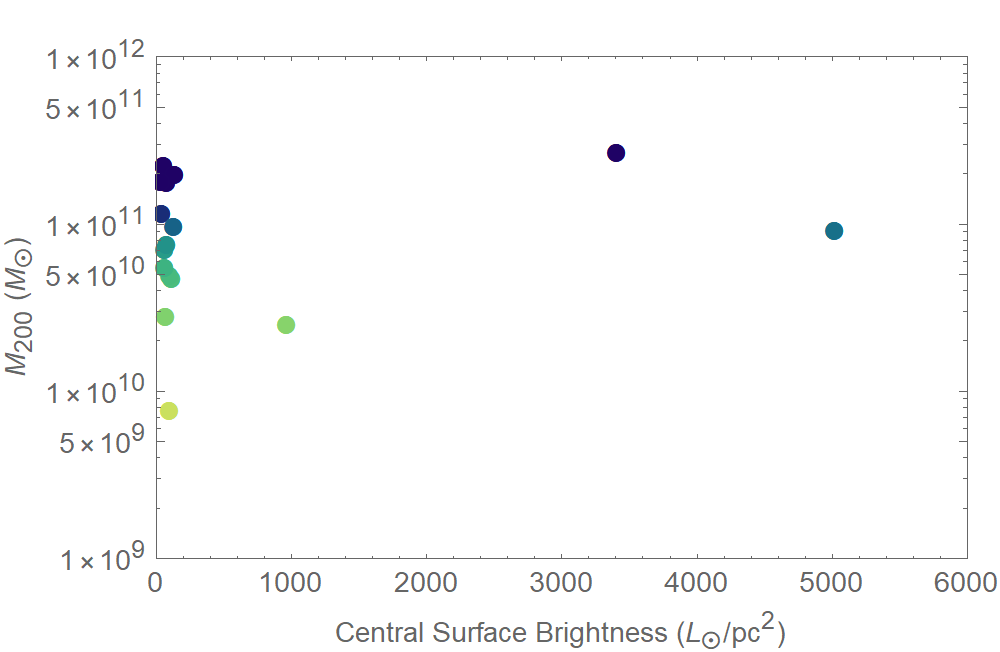}
\caption{Extracted virial masses $M_{200}$ in units of solar masses $M_{\odot}$ from sample of the 17 best-fitting galaxies in the SNFW model.}
\label{fig:GalaxyMassesSNFW}
\end{figure}

In Fig.\,\ref{fig:HistoMassesSNFW} a frequency distribution of $M_{200}$ is shown for the 17 best-fitting galaxies. Fig.\,\ref{fig:GalaxyMassesSNFW} depicts the distribution of these galaxies in the $M_{200}$ -- surface-brightness plane. The maximum of the smooth-kernel-distribution is at $M_{200}=(6.3\pm 3)\times 10^{10}\,M_{\odot}$. With Eq.\,\ref{Bohr} this implies a mean Bohr radius of  
\begin{align}
\label{maBurkert}
\nonumber r_{B,e}&=\frac{M_p^2}{(6.3\pm4) \times 10^{10}\,M_{\odot}\, ((0.72\pm 0.5) \times 10^{-23}\, {\rm eV})^2 }\\
&=(0.26\pm 0.1)\,{\rm kpc}\,.
\end{align}\nn
This value of $r_{B,e}$ is used in the Burkert-model analysis of Sec.\,\ref{BM} to extract the frequency distribution of $m_{a,e}$ via the frequency distribution of $M_{200}$.
\vspace{-6.8pt}
\vspace{15mm}

Fig.\,\ref{fig:GalaxyMassesSNFW} depicts the distribution of the sample of galaxies used in Fig.\,\ref{fig:AxionMassesSNFW} in the $M_{200}$ -- surface-brightness plane.\nn

\vspace{15pt}

\subsection{Analysis of RCs in the Burkert model}\label{BM}

Table\,4 depict the fits of the Burkert model to the 17 RCs used in the SNFW fits. Table\,5 and Table\,6 indicate that three out of these 17 RCs are fitted with a $\chi^2/{\rm d.o.f.}>1$. Therefore, we resort to a sample of 80 galaxies which fit with  $\chi^2/{\rm d.o.f.}<1$.\nn

\begin{figure}[H]
\centering
\includegraphics[width=7.5cm]{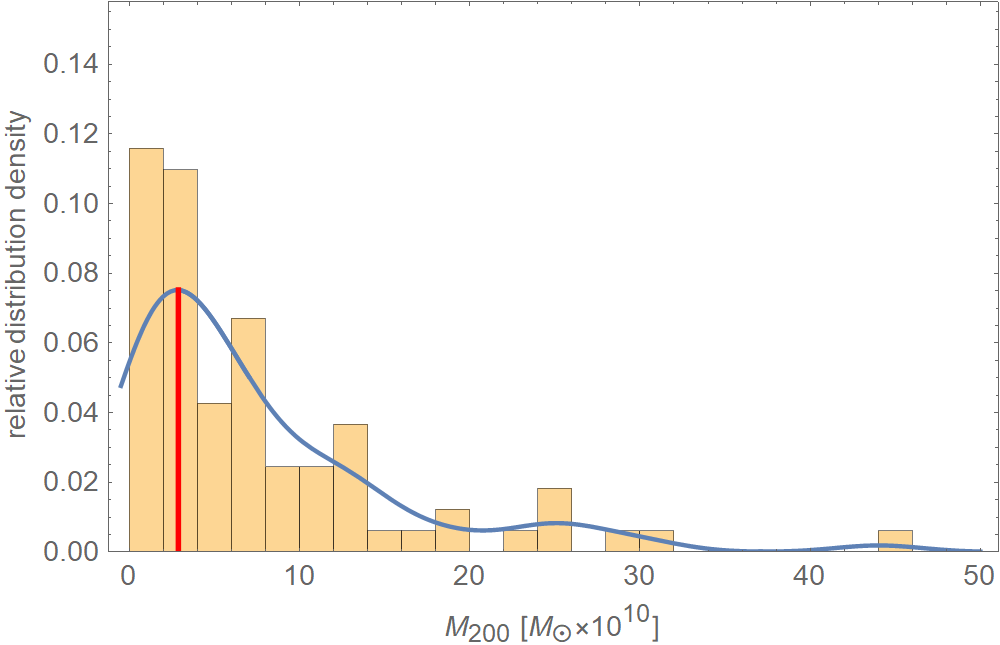}
\caption{
Frequency distribution of the virial mass $M_{200}$ in units of solar masses $M_{\odot}$ from the 80 best-fitting galaxies in the Burkert  model. The maximum %of the smooth-kernel-distribution (solid line) 
of the distribution is at $M_{200}=(2.9\pm4)\times 10^{10}\,M_{\odot}$.
}
\label{fig:M200Burkert}
\end{figure}

\begin{figure}[H]
\centering
\includegraphics[width=7.5cm]{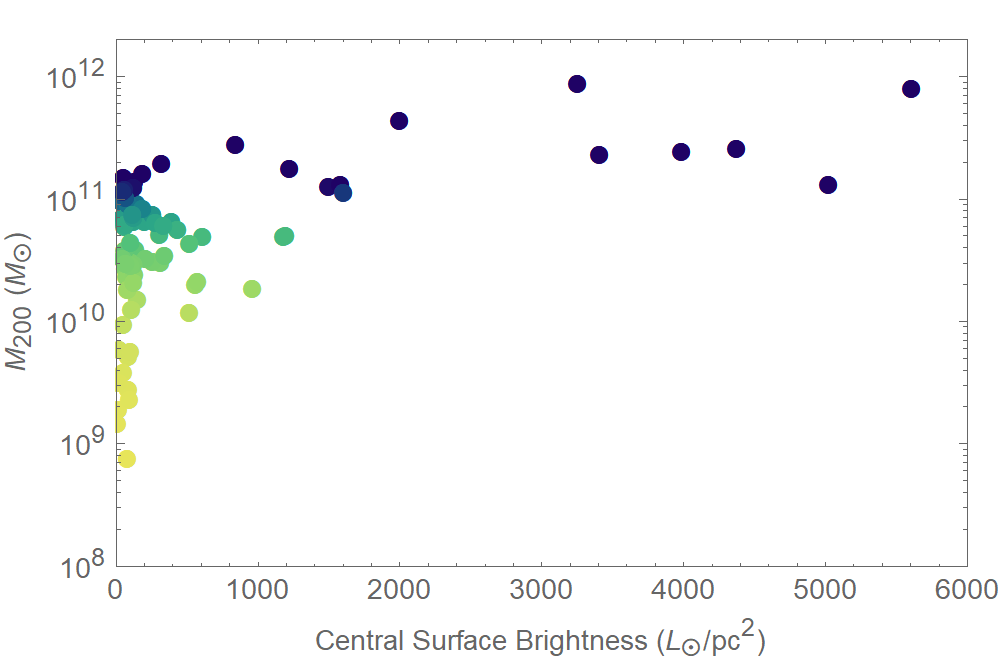}
\caption{Extracted virial masses $M_{200}$ in units of solar masses from Burkert-model fits to 80 RCs with $\chi^2/\text{d.o.f.}<1$ vs. the respective galaxy's central surface brightness in units of $L_{\odot}/\text{pc}^2$.}
\label{fig: M200  Burkert}
\end{figure}

Our strategy to demonstrate independence of the mean value of $m_{a,e}$ on the details of the two realistic models 
SNFW and Burkert is to also determine it from Eq.\,(\ref{Bohr}). To do this, we use the value of the gravitational Bohr radius $r_{B,e}$ 
in Eq.\,(\ref{maBurkert}) and the values of $M_{200}$ extracted from RC fits within an ensemble of 80 SPARC galaxies to the Burkert model. The results are characterised by Table\,5 and Table\,6,  Fig.\,\ref{fig:M200Burkert}, and  Fig.\,\ref{fig: M200  Burkert}.\nn

%\newpage

%-------------------- Burkert Tables 

\begin{figure*}
\textbf{Table 4.} Burkert-model fits to the 17 best fitting SNFW-model galaxies. The purple arrow indicates the value of $r_{0}$.\nn
\vspace{0.3cm}
\begin{minipage}{1\textwidth}
\centering
\hspace{-5mm}\includegraphics[width=17cm]{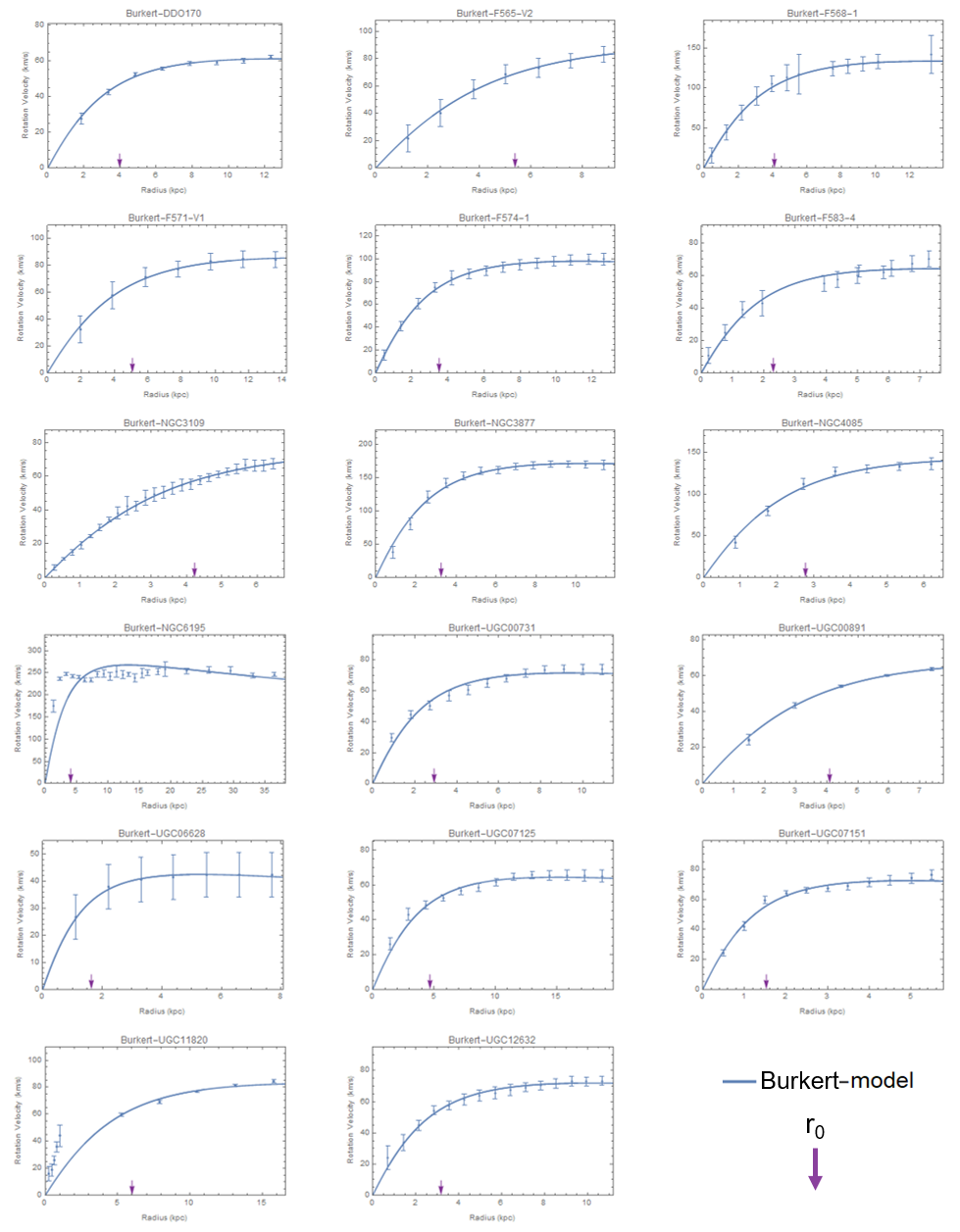}
\end{minipage}
\end{figure*}

\newpage

\end{multicols}{}

\nointerlineskip
    \begin{table}[H]
\caption*{{\bf Table 5.} Burkert model: Galaxy name, Hubble Type, $\chi^2/\text{d.o.f.}$, luminosity, axion mass $m_a$, $r_{200}$, virial mass $M_{200}$, core density $\rho_0$, and core radius $r_0$. The 17 galaxies used for the SNFW fit are highlighted in red.}\label{table4burkertfit}

\[  %a math mode needs to be introduced here
%{\small
\hspace{-1cm}\begin{array}{|l|c|c|c|c|c|c|c|c|}\hline
 \text{Galaxy} & 
\begin{array}{c}
 \text{Hub.} \\
 \text{Type} \\
\end{array}
 & 
\begin{array}{c}
 \text{Lum.} \\
 \left[L_{\odot}\right/\text{pc}^2] \\
\end{array}
 & 
\begin{array}{c}
 \text{$\chi $2 } \\
 \text{/d.o.f.} \\
\end{array}
 & 
\begin{array}{c}
 m_{a,e} \\
 \left.\text{[eV$\times $}10^{-23}\right] \\
\end{array}
 & 
\begin{array}{c}
 r_{200} \\
 \text{[kpc]} \\
\end{array}
 & 
\begin{array}{c}
 M_{200}  \\
 \left[M_{\odot}\times 10^{10}\right] \\
\end{array}
 & 
\begin{array}{c}
 \rho _0\times 10^7 \\
 \left[M_{\odot}\right/\text{kpc}^3] \\
\end{array}
 & 
\begin{array}{c}
 r_0 \\
 \text{[kpc]} \\
\end{array}
 \\
 \noalign{\global\arrayrulewidth=0.9 pt}
 \arrayrulecolor{spray}\hline\arrayrulecolor{black}
 \noalign{\global\arrayrulewidth=0.4 pt}

 %\setlength\arrayrulewidth{2pt}\arrayrulecolor{blue}
 
 %------------------------------ highlighted galaxies----- Burkert------------
 %\tikzmarkin{a}  \tikzmarkend{a} 
 %\tikzset{left offset=-0.02,right offset=0.02,disable rounded corners=true}

%\tikzmarkin{A}
%!{\color{spray}\vrule}
 \rowcolor{spray!10}\text{DDO170} & \text{Im} & 73.93 & 0.74 & 1.18\pm0.24 & 33.22\pm1.64 & 2.34\pm0.28 & 2.03\pm0.14 & 3.98\pm0.16 \\\hline
 \rowcolor{spray!10}\text{F565-V2} & \text{Im} & 40.26 & 0.04 & 0.68\pm0.14 & 47.44\pm3.11 & 7.06\pm1.03 & 2.39\pm0.16 & 5.39\pm0.29 \\\hline
 \rowcolor{spray!10}\text{F568-1} & \text{Sc} & 57.13 & 0.06 & 0.47\pm0.09 & 57.09\pm2.82 & 15.19\pm1.47 & 9.13\pm0.46 & 4.10\pm0.13 \\\hline
 \rowcolor{spray!10}\text{F571-V1} & \text{Sd} & 64.39 & 0.03 & 0.74\pm0.15 & 44.84\pm2.51 & 6.03\pm0.69 & 2.45\pm0.15 & 5.06\pm0.20 \\\hline
 \rowcolor{spray!10}\text{F574-1} & \text{Sd} & 128.48 & 0.10 & 0.70\pm0.14 & 43.92\pm1.74 & 6.62\pm0.45 & 6.69\pm0.26 & 3.51\pm0.08 \\\hline
 \rowcolor{spray!10}\text{F583-4} & \text{Sc} & 83.34 & 0.51 & 1.33\pm0.34 & 28.84\pm3.82 & 1.85\pm0.61 & 6.66\pm1.26 & 2.30\pm0.25 \\\hline
 \rowcolor{spray!10}\text{NGC3109} & \text{Sm} & 140.87 & 0.17 & 0.91\pm0.18 & 38.94\pm0.91 & 3.93\pm0.23 & 2.68\pm0.06 & 4.23\pm0.09 \\\hline
 \rowcolor{spray!10}\text{NGC3877} & \text{Sc} & 3410.59 & 0.46 & 0.38\pm0.08 & 63.43\pm3.48 & 23.22\pm3.28 & 23.82\pm1.88 & 3.26\pm0.15 \\\hline
 \rowcolor{spray!10}\text{NGC4085} & \text{Sc} & 5021.46 & 0.63 & 0.50\pm0.12 & 52.85\pm5.37 & 13.34\pm3.53 & 22.57\pm2.90 & 2.77\pm0.25 \\\hline
\rowcolor{spray!10} \text{NGC6195} & \text{Sb} & 174.11 & 50.97 & 0.21\pm0.06 & 91.02\pm16.07 & 74.49\pm32.29 & 37.13\pm9.18 & 4.07\pm0.56 \\\hline
 \rowcolor{spray!10}\text{UGC00731} & \text{Im} & 82.57 & 1.55 & 1.08\pm0.24 & 33.71\pm3.07 & 2.83\pm0.66 & 5.15\pm0.68 & 2.92\pm0.23 \\\hline
 \rowcolor{spray!10}\text{UGC00891} & \text{Sm} & 113.98 & 0.14 & 1.04\pm0.20 & 35.94\pm0.96 & 3.03\pm0.19 & 2.34\pm0.07 & 4.09\pm0.09 \\\hline
 \rowcolor{spray!10}\text{UGC06628} & \text{Sm} & 103.00 & 0.01 & 2.40\pm0.48 & 19.65\pm0.92 & 0.57\pm0.07 & 5.82\pm0.41 & 1.63\pm0.06 \\\hline
 \rowcolor{spray!10}\text{UGC07125} & \text{Sm} & 103.00 & 0.83 & 1.06\pm0.22 & 36.11\pm2.61 & 2.90\pm0.51 & 1.63\pm0.17 & 4.67\pm0.28 \\\hline
 \rowcolor{spray!10}\text{UGC07151} & \text{Scd} & 965.67 & 0.94 & 1.32\pm0.28 & 27.68\pm1.93 & 1.87\pm0.34 & 19.42\pm1.91 & 1.52\pm0.09 \\\hline
 \rowcolor{spray!10}\text{UGC11820} & \text{Sm} & 34.11 & 11.39 & 0.73\pm0.20 & 46.42\pm7.09 & 6.17\pm2.34 & 1.63\pm0.34 & 6.00\pm0.80 \\\hline
 \rowcolor{spray!10}\text{UGC12632}  & \text{Sm} & 66.81 & 0.22 & 1.05\pm0.21 & 34.69\pm1.37 & 3.01\pm0.30 & 4.40\pm0.25 & 3.18\pm0.10 %\tikzmarkend{A} 
 \\
 \noalign{\global\arrayrulewidth=0.9 pt}
 \arrayrulecolor{spray}\hline\arrayrulecolor{black}
 \noalign{\global\arrayrulewidth=0.4 pt}

 %------------------------------------------------
 \text{CamB} & \text{Im} & 66.20 & 0.02 & 1.10\pm0.31 & 36.66\pm6.71 & 2.74\pm1.15 & 0.91\pm0.04 & 5.81\pm1.06 \\\hline
 \text{D512-2} & \text{Im} & 93.94 & 0.33 & 2.52\pm0.76 & 19.25\pm3.55 & 0.52\pm0.24 & 4.68\pm0.99 & 1.73\pm0.29 \\\hline
 \text{D564-8} & \text{Im} & 21.13 & 0.02 & 4.15\pm0.82 & 14.37\pm0.50 & 0.19\pm0.02 & 2.10\pm0.08 & 1.70\pm0.05 \\\hline
 \text{DDO064} & \text{Im} & 151.65 & 0.40 & 1.47\pm0.45 & 27.02\pm4.99 & 1.51\pm0.71 & 6.87\pm1.10 & 2.12\pm0.37 \\\hline
 \text{F563-1} & \text{Sm} & 41.77 & 0.54 & 0.59\pm0.13 & 50.11\pm4.84 & 9.55\pm2.26 & 5.90\pm0.83 & 4.16\pm0.32 \\\hline
 \text{F563-V2} & \text{Im} & 146.16 & 0.15 & 0.60\pm0.13 & 47.48\pm3.69 & 9.24\pm1.70 & 14.65\pm1.42 & 2.89\pm0.18 \\\hline
 \text{F567-2} & \text{Sm} & 46.65 & 0.25 & 1.64\pm0.56 & 26.26\pm6.02 & 1.23\pm0.69 & 2.66\pm0.82 & 2.88\pm0.56 \\\hline
 \text{F568-3} & \text{Sd} & 132.08 & 0.80 & 0.49\pm0.10 & 58.93\pm5.29 & 13.95\pm2.73 & 2.76\pm0.23 & 6.39\pm0.47 \\\hline
 \text{F568-V1} & \text{Sd} & 90.54 & 0.05 & 0.60\pm0.12 & 47.54\pm2.15 & 9.09\pm0.82 & 12.07\pm0.57 & 3.10\pm0.09 \\\hline
 \text{F579-V1} & \text{Sc} & 201.76 & 0.47 & 0.71\pm0.17 & 41.30\pm4.54 & 6.59\pm1.75 & 25.80\pm4.07 & 2.08\pm0.17 \\\hline
 \text{F583-1} & \text{Sm} & 60.93 & 0.11 & 0.73\pm0.14 & 45.23\pm1.46 & 6.23\pm0.43 & 2.73\pm0.08 & 4.90\pm0.13 \\\hline
 \text{KK98-251} & \text{Im} & 52.10 & 0.57 & 1.86\pm0.46 & 24.45\pm3.06 & 0.95\pm0.29 & 2.25\pm0.22 & 2.83\pm0.34 \\\hline
 \text{NGC0024} & \text{Sc} & 1182.58 & 0.61 & 0.81\pm0.16 & 36.96\pm1.41 & 5.03\pm0.51 & 51.79\pm3.18 & 1.46\pm0.05 \\\hline
 \text{NGC0055} & \text{Sm} & 391.59 & 0.32 & 0.70\pm0.14 & 46.22\pm1.19 & 6.62\pm0.42 & 2.80\pm0.08 & 4.94\pm0.12 \\\hline
 \text{NGC0100} & \text{Scd} & 1193.52 & 0.10 & 0.81\pm0.16 & 40.65\pm1.10 & 5.04\pm0.33 & 5.73\pm0.18 & 3.40\pm0.08 \\\hline
 \text{NGC0300} & \text{Sd} & 437.35 & 0.74 & 0.76\pm0.15 & 41.89\pm1.63 & 5.71\pm0.58 & 7.45\pm0.43 & 3.20\pm0.11 \\\hline
 \text{NGC2366} & \text{Im} & 113.98 & 0.69 & 1.61\pm0.32 & 25.86\pm1.22 & 1.27\pm0.15 & 4.98\pm0.29 & 2.27\pm0.10 \\\hline
 \text{NGC2915} & \text{BCD} & 313.93 & 0.30 & 1.04\pm0.22 & 32.66\pm2.31 & 3.03\pm0.58 & 17.25\pm2.02 & 1.87\pm0.11 \\\hline
 \text{NGC2976} & \text{Sc} & 1502.55 & 0.62 & 0.51\pm0.12 & 52.34\pm5.22 & 12.77\pm3.27 & 20.73\pm1.02 & 2.82\pm0.28 \\\hline
 \text{NGC3917} & \text{Scd} & 1226.96 & 0.87 & 0.43\pm0.09 & 61.13\pm2.97 & 18.06\pm2.19 & 8.28\pm0.54 & 4.51\pm0.18 \\\hline
 \text{NGC3949} & \text{Sbc} & 185.71 & 0.85 & 0.45\pm0.11 & 54.76\pm6.26 & 16.38\pm4.96 & 50.95\pm7.60 & 2.18\pm0.22 \\\hline
 \text{NGC3953} & \text{Sbc} & 1999.08 & 0.20 & 0.27\pm0.05 & 77.09\pm3.05 & 44.03\pm4.44 & 37.19\pm2.18 & 3.41\pm0.11 \\\hline
 \text{NGC3972} & \text{Sbc} & 1587.93 & 0.51 & 0.50\pm0.10 & 53.96\pm3.61 & 13.22\pm2.25 & 13.02\pm1.14 & 3.41\pm0.20 \\\hline
 \text{NGC3992} & \text{Sbc} & 3257.09 & 0.95 & 0.19\pm0.04 & 99.24\pm8.40 & 88.76\pm19.86 & 23.20\pm3.34 & 5.15\pm0.34 \\\hline
 \text{NGC4068} & \text{Im} & 261.11 & 0.07 & 1.03\pm0.28 & 35.59\pm5.60 & 3.09\pm1.19 & 3.22\pm0.24 & 3.63\pm0.56 \\\hline
 \text{NGC4088} & \text{Sbc} & 3988.69 & 0.65 & 0.36\pm0.08 & 64.61\pm4.96 & 24.74\pm5.03 & 25.43\pm3.19 & 3.25\pm0.20 \\\hline
 \text{NGC4157} & \text{Sb} & 23813.87 & 0.85 & 0.33\pm0.07 & 68.59\pm5.23 & 30.54\pm6.20 & 32.73\pm4.20 & 3.17\pm0.19 \\\hline
 \text{NGC4217} & \text{Sb} & 4373.51 & 0.55 & 0.36\pm0.07 & 64.62\pm2.90 & 25.96\pm3.00 & 37.45\pm2.57 & 2.85\pm0.10 \\\hline
 \text{NGC4389} & \text{Sbc} & 322.72 & 0.15 & 0.41\pm0.10 & 63.93\pm7.79 & 19.69\pm5.94 & 5.91\pm0.55 & 5.29\pm0.61 \\\hline
 \text{NGC4559} & \text{Scd} & 1602.62 & 0.76 & 0.54\pm0.11 & 52.00\pm2.59 & 11.35\pm1.47 & 9.78\pm0.78 & 3.62\pm0.15 \\\hline
 \text{UGC00634} & \text{Sm} & 126.13 & 0.87 & 0.51\pm0.11 & 57.24\pm4.75 & 12.59\pm2.54 & 2.70\pm0.31 & 6.22\pm0.43 \\\hline
 \text{UGC01230} & \text{Sm} & 69.32 & 0.19 & 0.56\pm0.12 & 51.30\pm3.78 & 10.35\pm1.81 & 6.27\pm0.67 & 4.18\pm0.23 \\\hline
\end{array}
%}
\]

\end{table}

%########################## second part ##############
%\newpage

\nointerlineskip
    \begin{table}[H]
\caption*{{\bf Table 6.} Burkert model with $\chi^2/{\rm d.o.f.}<1$: Galaxy name, Hubble Type, $\chi^2/\text{d.o.f.}$, luminosity, axion mass $m_a$, $r_{200}$, virial mass $M_{200}$, core density $\rho_0$, and core radius $r_0$ (ordered as in the SPARC library).}\label{table5burkertfit}

\[
%{\small
\hspace{-1cm}\begin{array}{|l|c|c|c|c|c|c|c|c|}\hline
 \text{Galaxy} & 
\begin{array}{c}
 \text{Hub.} \\
 \text{Type} \\
\end{array}
 & 
\begin{array}{c}
 \text{Lum.} \\
 \left[L_{\odot}\right/\text{pc}^2] \\
\end{array}
 & 
\begin{array}{c}
 \text{$\chi $2 } \\
 \text{/d.o.f.} \\
\end{array}
 & 
\begin{array}{c}
 m_{a,e} \\
 \left.\text{[eV$\times $}10^{-23}\right] \\
\end{array}
 & 
\begin{array}{c}
 r_{200} \\
 \text{[kpc]} \\
\end{array}
 & 
\begin{array}{c}
 M_{200}  \\
 \left[M_{\odot}\times 10^{10}\right] \\
\end{array}
 & 
\begin{array}{c}
 \rho _0\times 10^7 \\
 \left[M_{\odot}\right/\text{kpc}^3] \\
\end{array}
 & 
\begin{array}{c}
 r_0 \\
 \text{[kpc]} \\
\end{array}
 \\\hline
 \text{UGC01281} & \text{Sdm} & 135.78 & 0.17 & 1.16\pm0.24 & 32.78\pm1.89 & 2.45\pm0.35 & 3.56\pm0.20 & 3.23\pm0.17 \\\hline
 \text{UGC02023} & \text{Im} & 121.57 & 0.03 & 0.36\pm0.16 & 73.23\pm24.56 & 24.82\pm19.87 & 1.92\pm0.21 & 8.95\pm2.97 \\\hline
 \text{UGC04305} & \text{Im} & 88.07 & 0.76 & 3.42\pm0.90 & 15.39\pm2.06 & 0.28\pm0.10 & 7.20\pm1.55 & 1.19\pm0.14 \\\hline
 \text{UGC04325} & \text{Sm} & 213.22 & 0.26 & 1.00\pm0.20 & 32.71\pm1.31 & 3.29\pm0.34 & 31.75\pm1.83 & 1.53\pm0.05 \\\hline
 \text{UGC04483} & \text{Im} & 82.57 & 0.26 & 6.58\pm1.48 & 9.51\pm0.84 & 0.08\pm0.02 & 19.32\pm2.23 & 0.52\pm0.04 \\\hline
 \text{UGC04499} & \text{Sdm} & 127.30 & 0.09 & 1.05\pm0.21 & 34.00\pm1.01 & 2.96\pm0.21 & 5.97\pm0.24 & 2.80\pm0.07 \\\hline
 \text{UGC05005} & \text{Im} & 65.59 & 0.06 & 0.52\pm0.10 & 59.27\pm2.86 & 12.09\pm1.21 & 1.10\pm0.06 & 8.85\pm0.32 \\\hline
 \text{UGC05414} & \text{Im} & 127.30 & 0.11 & 1.25\pm0.25 & 30.31\pm1.42 & 2.09\pm0.24 & 5.90\pm0.30 & 2.51\pm0.11 \\\hline
 \text{UGC05750} & \text{Sdm} & 124.98 & 0.08 & 0.68\pm0.14 & 49.77\pm3.23 & 7.13\pm0.97 & 1.07\pm0.06 & 7.50\pm0.40 \\\hline
 \text{UGC05829} & \text{Im} & 63.22 & 0.33 & 0.93\pm0.23 & 38.90\pm4.80 & 3.81\pm1.16 & 2.21\pm0.28 & 4.52\pm0.52 \\\hline
 \text{UGC05918} & \text{Im} & 24.94 & 0.06 & 2.35\pm0.48 & 19.90\pm1.15 & 0.59\pm0.09 & 6.01\pm0.47 & 1.64\pm0.08 \\\hline
 \text{UGC05999} & \text{Im} & 51.62 & 0.35 & 0.53\pm0.12 & 57.19\pm5.87 & 11.71\pm2.81 & 1.72\pm0.22 & 7.29\pm0.63 \\\hline
 \text{UGC06399} & \text{Sm} & 311.05 & 0.03 & 0.80\pm0.16 & 40.82\pm1.20 & 5.16\pm0.36 & 6.20\pm0.22 & 3.32\pm0.08 \\\hline
 \text{UGC06446} & \text{Sd} & 86.46 & 0.78 & 1.05\pm0.23 & 32.95\pm2.75 & 3.01\pm0.66 & 13.01\pm1.64 & 2.08\pm0.15 \\\hline
 \text{UGC06667} & \text{Scd} & 614.94 & 0.11 & 0.81\pm0.16 & 40.55\pm1.30 & 5.03\pm0.39 & 5.97\pm0.23 & 3.34\pm0.09 \\\hline
 \text{UGC06917} & \text{Sm} & 261.11 & 0.35 & 0.66\pm0.13 & 45.28\pm2.24 & 7.48\pm0.92 & 9.48\pm0.64 & 3.19\pm0.13 \\\hline
 \text{UGC06923} & \text{Im} & 347.40 & 0.40 & 0.97\pm0.26 & 34.69\pm4.92 & 3.46\pm1.28 & 11.64\pm2.17 & 2.28\pm0.29 \\\hline
 \text{UGC06930} & \text{Sd} & 189.16 & 0.21 & 0.62\pm0.13 & 47.70\pm3.06 & 8.45\pm1.38 & 7.41\pm0.72 & 3.65\pm0.19 \\\hline
 \text{UGC06983} & \text{Scd} & 121.57 & 0.68 & 0.66\pm0.14 & 45.05\pm3.30 & 7.54\pm1.42 & 11.27\pm1.24 & 2.99\pm0.18 \\\hline
 \text{UGC07089} & \text{Sdm} & 520.99 & 0.22 & 0.86\pm0.18 & 40.29\pm2.82 & 4.41\pm0.75 & 2.82\pm0.23 & 4.31\pm0.27 \\\hline
 \text{UGC07261} & \text{Sdm} & 566.02 & 0.33 & 1.28\pm0.31 & 28.32\pm3.03 & 2.00\pm0.57 & 18.97\pm3.11 & 1.57\pm0.14 \\\hline
 \text{UGC07323} & \text{Sdm} & 283.68 & 0.80 & 0.71\pm0.17 & 44.67\pm5.34 & 6.55\pm1.98 & 5.06\pm0.62 & 3.90\pm0.43 \\\hline
 \text{UGC07524} & \text{Sm} & 106.86 & 0.57 & 0.86\pm0.17 & 39.92\pm1.58 & 4.45\pm0.44 & 3.66\pm0.19 & 3.89\pm0.13 \\\hline
 \text{UGC07559} & \text{Im} & 55.06 & 0.03 & 2.93\pm0.60 & 17.50\pm0.93 & 0.38\pm0.05 & 4.15\pm0.23 & 1.63\pm0.08 \\\hline
 \text{UGC07577} & \text{Im} & 54.55 & 0.04 & 3.14\pm1.07 & 18.08\pm4.38 & 0.33\pm0.19 & 1.02\pm0.12 & 2.76\pm0.66 \\\hline
 \text{UGC07603} & \text{Sd} & 520.99 & 0.22 & 1.66\pm0.33 & 23.70\pm1.00 & 1.19\pm0.13 & 21.70\pm1.30 & 1.26\pm0.05 \\\hline
 \text{UGC07608} & \text{Im} & 46.65 & 0.07 & 1.00\pm0.22 & 35.39\pm2.78 & 3.27\pm0.65 & 5.23\pm0.44 & 3.05\pm0.22 \\\hline
 \text{UGC07866} & \text{Im} & 97.46 & 0.10 & 3.78\pm0.92 & 14.34\pm1.63 & 0.23\pm0.07 & 7.72\pm1.11 & 1.08\pm0.11 \\\hline
 \text{UGC08490} & \text{Sm} & 576.54 & 0.95 & 1.24\pm0.25 & 28.16\pm1.51 & 2.14\pm0.31 & 38.07\pm3.44 & 1.23\pm0.05 \\\hline
 \text{UGC08837} & \text{Im} & 77.42 & 0.29 & 0.54\pm0.18 & 57.39\pm13.51 & 11.12\pm6.14 & 1.26\pm0.11 & 8.13\pm1.89 \\\hline
 \text{UGC09037} & \text{Scd} & 841.07 & 0.61 & 0.34\pm0.07 & 72.32\pm3.54 & 28.35\pm2.75 & 5.11\pm0.28 & 6.33\pm0.21 \\\hline
 \text{UGC09992} & \text{Im} & 73.25 & 0.24 & 3.68\pm1.34 & 14.37\pm3.31 & 0.24\pm0.15 & 10.76\pm3.86 & 0.97\pm0.19 \\\hline
 \text{UGC11557} & \text{Sdm} & 337.93 & 0.28 & 0.73\pm0.17 & 44.75\pm4.74 & 6.16\pm1.60 & 3.16\pm0.40 & 4.60\pm0.43 \\\hline
 \text{UGC12506} & \text{Scd} & 5608.28 & 0.69 & 0.20\pm0.04 & 96.54\pm6.95 & 80.52\pm12.65 & 17.80\pm1.77 & 5.53\pm0.26 \\\hline
 \text{UGCA281} & \text{BCD} & 12.05 & 0.34 & 4.71\pm1.08 & 11.71\pm1.10 & 0.15\pm0.04 & 27.12\pm2.79 & 0.58\pm0.05 \\\hline
\end{array}
%} %for the \small environment 
\]
  \end{table}

\begin{multicols}{2}

\begin{figure}[H]\centering
\includegraphics[width=7.5cm]{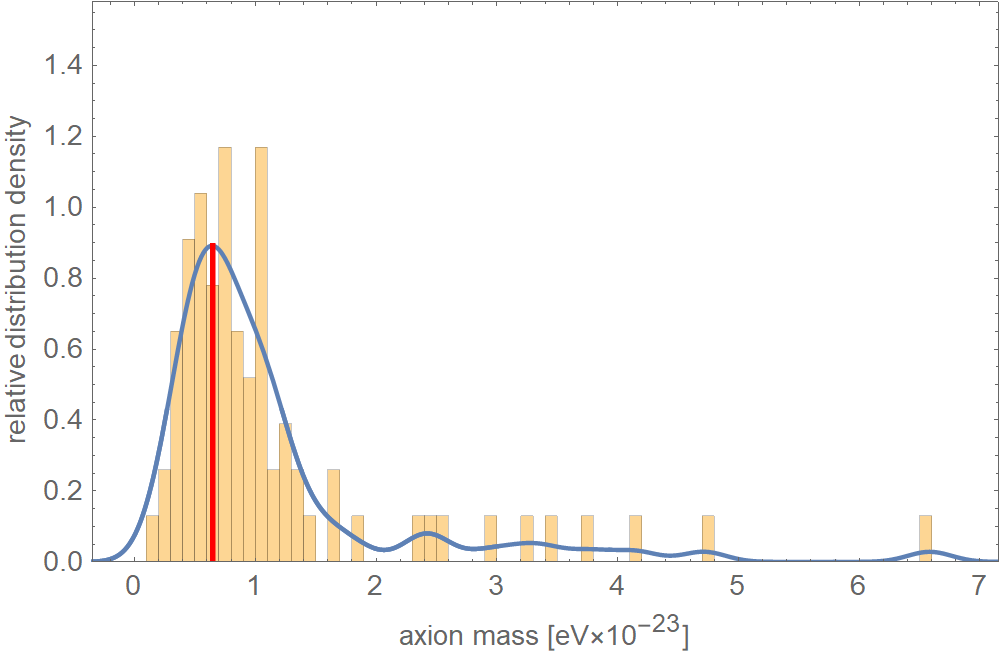}
\caption{Frequency distribution of 80 axion masses $m_{a,e}$, extracted from the Burkert-model fits of $M_{200}$ to the RCs of galaxies with a $\chi^2/\text{d.o.f.}<1$. The maximum of the smooth-kernel distribution (solid, blue line) is $m_{a,e}=(0.65\pm0.4)\times10^{-23}$\,eV (red line).}
\label{fig:AxionMassesBurkert}
\end{figure}

This yields a frequency distribution of $m_{a,e}$ shown in Fig.\,\ref{fig:AxionMassesBurkert}. Obviously, the maximum of the smooth-kernel distribution, $m_{a,e}=(0.65\pm 0.4)\times 10^{-23}\,$eV, is compatible with that in the SNFW model $m_{a,e}=(0.72\pm 0.5)\times 10^{-23}\,{\rm eV}$. Notice how $M_{200}$ clusters around the value $M_{200}\sim 5\times 10^{10}\,M_{\odot}$.\nn

In our treatment Sec.\,\ref{Sec5} of cosmological and astrophysical implications we appeal to the mean value of $m_{a,e}$-extractions in the SNFW and the Burkert model as 
\vspace{-2mm}
\begin{equation}
\label{maBurkertSNFW}
m_{a,e}=0.675\times 10^{-23}\,{\rm} {\rm eV}\,.
\end{equation}

%-----
%\vspace{-1.2cm}

\section{Galactic central regions and the dark sector of the Universe}\label{Sec4}

Interpreting the dark-matter structure of a typical low-surface-brightness galaxy as an e-lump, we have $r_{B,e}= 0.26\,$kpc from the SNFW model, see Sec.\,\ref{SNFW}. Therefore, the value of $\kappa$ in Eq.\,(\ref{CvB}) is
\vspace{3pt}
\begin{equation}
\label{kappavalue}
\kappa=314\,.
\end{equation}
With $m_{a,e}=0.675\times 10^{-23}\,$eV  Eq.\,(\ref{maxxaxion}) yields 
\vspace{3pt}
\begin{equation}
\label{YMeextr}
\Lambda_e=287\,\mbox{eV}\,.
\end{equation}
This is by only a factor 15 smaller than the scale $\Lambda_e=m_e/118.6$ ($m_e=511\,$keV the mass of the electron) of an SU(2) Yang-Mills theory proposed in \cite{Hofmann:2017lmu} to originate the electron's mass in terms of a fuzzy ball of deconfining phase. There the deconfining region is immersed into the confining phase and formed by the selfintersection of a center-vortex loop. Considering an undistorted Yang-Mills theory for simplicity\footnote{The chiral dynamics at the Planck scale, which produces the axion field, to some extent resolves the ground states of Yang-Mills theories: axions become massive by virtue of the anomaly because of this very resolution of topological charge density.}, the factor of 15 could be explained by a stronger screening of topological charge density -- the origin of the axial anomaly -- in the confining ground state, composed of round, pointlike center-vortex loops, versus the deconfining thermal ground state, made of densely packed, spatially extended (anti)caloron centers subject to overlapping peripheries \cite{bookHofmann}. The factor of 15 so far is a purely phenomenological result (it could be expected to be O(100) or higher) which is plausible qualitatively because of the reduced topological charge density in the confining phase where overlapping magnetic monopoles and antimonopoles, aligned within hardly resolved 
center vortices, are the topological charge carriers. The complex interplay between the would-be Goldstone nature of the axion, as prescribed by fermion interaction at the Planck scale, and the topological charge density of an SU(2) Yang-Mills theory deeply 
in its confining phase is anything but understood quantitatively so far. One may hope that simulations of the axion potential in a center-vortex model of the confining phase, such a proposed in \cite{Engelhardt:2000wc}, will yield more quantitative insights in the future. 
\nn
The link between the masses of the three species of ultralight axions, whose fuzzy condensates form lumps of typical masses $M_e$, $M_\mu$, and $M_\tau$, with the three lepton families via the Planck-scale originated axial anomaly within confining phases of SU(2) Yang-Mills theories is compelling. In particular, $M_e=M_{200}$ can be determined by mild modelling of direct observation, as done in Sec.\,\ref{Sec3}, while $M_\mu$ and $M_\tau$ are predicted by an appeal to Eqs.\,(\ref{lumpmassratios}). Such a scenario allows to address two questions: (i) the implication of a given lump's selfgravity for its stability and (ii) the cosmological origin of a given species of isolated lumps.\nn

%---

Before we discuss question (i) we would like to provide a 
thermodynamical argument, based on our knowledge gained about axion and lump masses in terms of Yang-Mills scales and the Planck mass, why Planck-scale axions associated with the lepton families always occur in the form of fuzzy or homogeneous condensates. Namely, the Yang-Mills scales $\Lambda_e$, $\Lambda_\mu=\Lambda_e\, m_\mu/m_e$, and $\Lambda_\tau=\Lambda_e\, m_\tau/m_e$ together with Eqs.\,(\ref{maxxaxion}), (\ref{YMeextr}), yield axion masses as 
\begin{align}
\label{axionmassemutau}
\nonumber&m_{a,e}\sim 0.675\times10^{-23}\,\mbox{eV} \,,\\ 
&m_{a,\mu}\sim 2.89\times10^{-19}\,\mbox{eV} \,,\\
\nonumber&m_{a,\tau}\sim 8.17\times 10^{-17}\,\mbox{eV}\,.
\end{align}\nn
The critical temperature $T_c$ for the Bose-Einstein condensation of a quantum gas of free bosons of mass $m_a$ and (mean) number density $n_a\sim M/(m_a \frac43\pi r_B^3)$ is given as 
 \begin{equation}
\label{condfrBos}
T_c=\frac{2\pi}{m_a}\left(\frac{n_a}{\zeta(3/2)}\right)^{2/3}\,.
\end{equation}\nn
We conclude from Eqs. (\ref{lumpmassratios}),\, (\ref{Bohrradiusfund}),\,(\ref{axionmassemutau}) and (\ref{condfrBos}) that
\begin{align}
\label{critTs}
\nonumber&T_{c,e}\sim 9.7\times 10^{30}\,{\rm GeV}\,,\\ 
&T_{c,\mu}\sim 7.7\times 10^{39}\,{\rm GeV}\,,\\
\nonumber&T_{c,\tau}\sim 6.1\times 10^{42}\,{\rm GeV}\,. 
\end{align}
All three critical temperatures are comfortably larger than the Planck mass $M_P=1.22\times 10^{19}\,$GeV such that throughout the Universe's expansion history and modulo depercolation, which generates a nonthermal halo of particles correlated on the de Broglie wave length around a condensate core, the Bose-condensed state of e-, $\mu$-, and $\tau$-axions is guaranteed and consistent with $\xi\ll 1$, compare with Eq.\,(\ref{xidef}).\nn

\begin{figure}[H]\centering
\includegraphics[width=7.5cm]{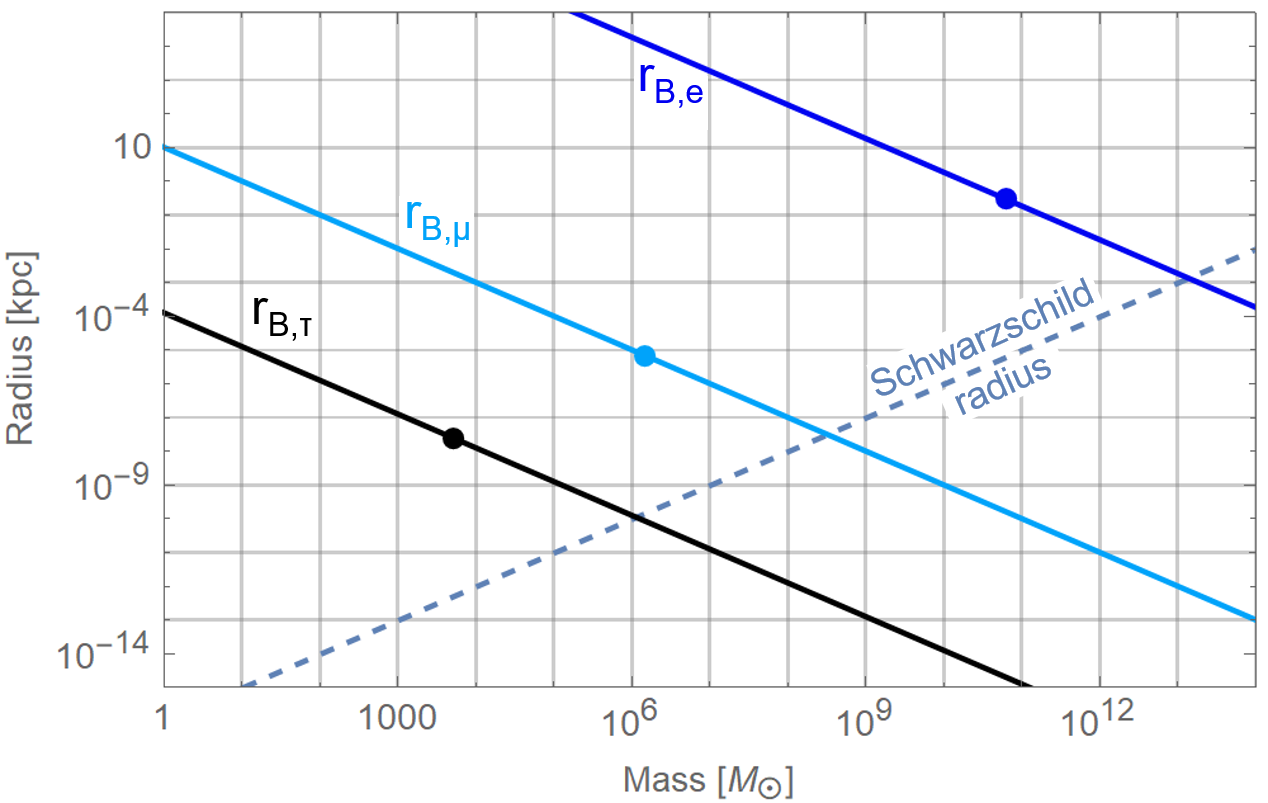}
\caption{Schwarzschild radius (dashed line) and gravitational Bohr radii (solid lines; dark blue: e-lump, turquoise: $\mu$-lump, and black: $\tau$-lump) as functions of lump mass in units of solar mass $M_\odot$. The dots indicate lump masses which derive from the typical mass of an isolated e-lump $M_e=6.3\times 10^{10}\,M_{\odot}$ suggested by the analysis of the RCs of low-surface-brightness galaxies performed in Sec.\,\ref{Sec3}.}
\label{fig:SchwarzschieldradiusVSBohrradii}
\end{figure}

We now turn back to question (i). Explicit lump masses can be obtained from Eqs.\,(\ref{lumpmassratios}) based on the typical mass $M_e=6.3\times 10^{10}\,M_{\odot}$ of an e-lump. One has
\begin{align}
\label{lumpmassemutau}
&M_\mu=1.5\times 10^{6}\,M_\odot\,,\nonumber\\ 
&M_\tau=5.2\times 10^{3}\,M_\odot\,.
\end{align}
For the computation of the respective gravitational Bohr radii according to Eq.\,(\ref{Bohr}) both quantities, axion mass $m_{a,i}$ and lump mass $M_i$, are required. To judge the gravitational stability of a given isolated and unmerged lump throughout its evolution a comparison between the typical Bohr radius $r_{B,i}$ and the typical Schwarzschild radius $r_{\rm SD,i}$, defined as 
 \begin{equation}
\label{Schwarzschildradius}
r_{\rm SD,i}\equiv\frac{2M_i}{M_P^2}\,,
\end{equation}\nn
is in order. Using $M_e=6.3\times 10^{10}\,M_{\odot}$, Fig.\,\ref{fig:SchwarzschieldradiusVSBohrradii} indicates the implied values of the Bohr radii 
$r_{B,e}$, $r_{B,\mu}$, and $r_{B,\tau}$ by dots on the curves of all possible Bohr radii as functions of their lump masses when keeping the axion mass $m_{a,i}$ fixed. Notice that for all three cases, e-lumps, $\mu$-lumps, and $\tau$-lumps, typical Bohr radii are considerably larger than their Schwarzschild radii. Indeed, 
from Eqs.\,(\ref{maxxaxion}), (\ref{MLambda}), and (\ref{Schwarzschildradius}) it follows that
 \begin{equation}\label{ratioBSD}
\frac{r_B}{r_{\rm SD}}=\frac12 \kappa^2\,.
\end{equation}\nn
With $\kappa=314$ we have $r_B/r_{\rm SD}=4.92\times 10^4$. An adiabatic pursuit of the solid lines in Fig.\,\ref{fig:SchwarzschieldradiusVSBohrradii} down to their intersections with the dashed line reveals that an  increase of lump mass by a factor $\sim 222$ is required to reach the critical mass for black-hole formation. While this is unlikely to occur through mergers of e-lumps within their peers it is conceivable for merging $\mu$- and $\tau$-lumps, see below.\nn

The mean mass density of a lump scales with the fourth power of the Yang-Mills scale, see Eqs.\,(\ref{densmean}), (\ref{MLambda}), and (\ref{Bohrradiusfund}). With the hierarchies in Yang-Mills scales $\Lambda_\tau/\Lambda_\mu\sim 17$ or $\Lambda_\mu/\Lambda_e\sim 200$ it is conceivable that sufficiently large number of lumps of a higher Yang-Mills scale, embedded into a lump of a lower scale, catalyse the latter's gravitational compaction to the point of collapse, see, however, discussion in Sec.\,5.1.    
\nn
With Eq.\,(\ref{lumpmassratios}) we have $M_\mu/M_e \sim 2.3\times 10^{-5}$ such that a dark mass of the selfgravitating dark-matter disk of the Milky Way, exhibiting a radial scale of $(7.5\cdots 8.85)\,$kpc and a mass of $M_{\rm MW}=(2\cdots 3)\times 10^{11}\,M_{\odot}$ \cite{Kalberla:2007sr}, would contain a {\sl few} previously isolated but now merged e-lumps. This implies with Eqs.\,(\ref{lumpmassratios}) a $\mu$-lump mass of 
 \begin{equation}
\label{compobjMW}
M_\mu=(4.7\cdots 7)\times10^{6}\,M_\odot\,.
\end{equation}
In \cite{Kalberla:2007sr} the mass of the dark halo of the Milky Way, which is virialised up to $r\sim 350$\,kpc, is determined as $1.8\times 10^{12}\,M_\odot$. In addition to the halo and the disk, there is a ringlike dark-matter structure within (13\,$\cdots$ 18.5)\,kpc of mass $(2.2\cdots 2.8)\times 10^{10}\,M_\odot$. Since these structures probably are, judged within the here-discussed framework, due to contaminations of a seeding e-lump by the accretion of $\tau$- and $\mu$-lumps we ignore them in what follows. In any case, a virialised dark-matter halo of 350\,kpc radial extent easily accomodates the dark mass ratio $\sim 0.1$ between the selfgravitating dark-matter disk and the dark halo in terms of accreted $\tau$- and $\mu$-lumps. 

Interestingly, the lower mass bound of Eq.\,(\ref{compobjMW}) is contained in the mass range $(4.5\pm 0.4)\times 10^6\,M_\odot$ \cite{Ghez2008} or $(4.31\pm 0.36)\times 10^6\,M_\odot$ \cite{Gillessen:2008qv} of the central compact object extracted from orbit analysis of S-stars.\nn

\begin{figure}[H]\centering
\includegraphics[width=8cm]{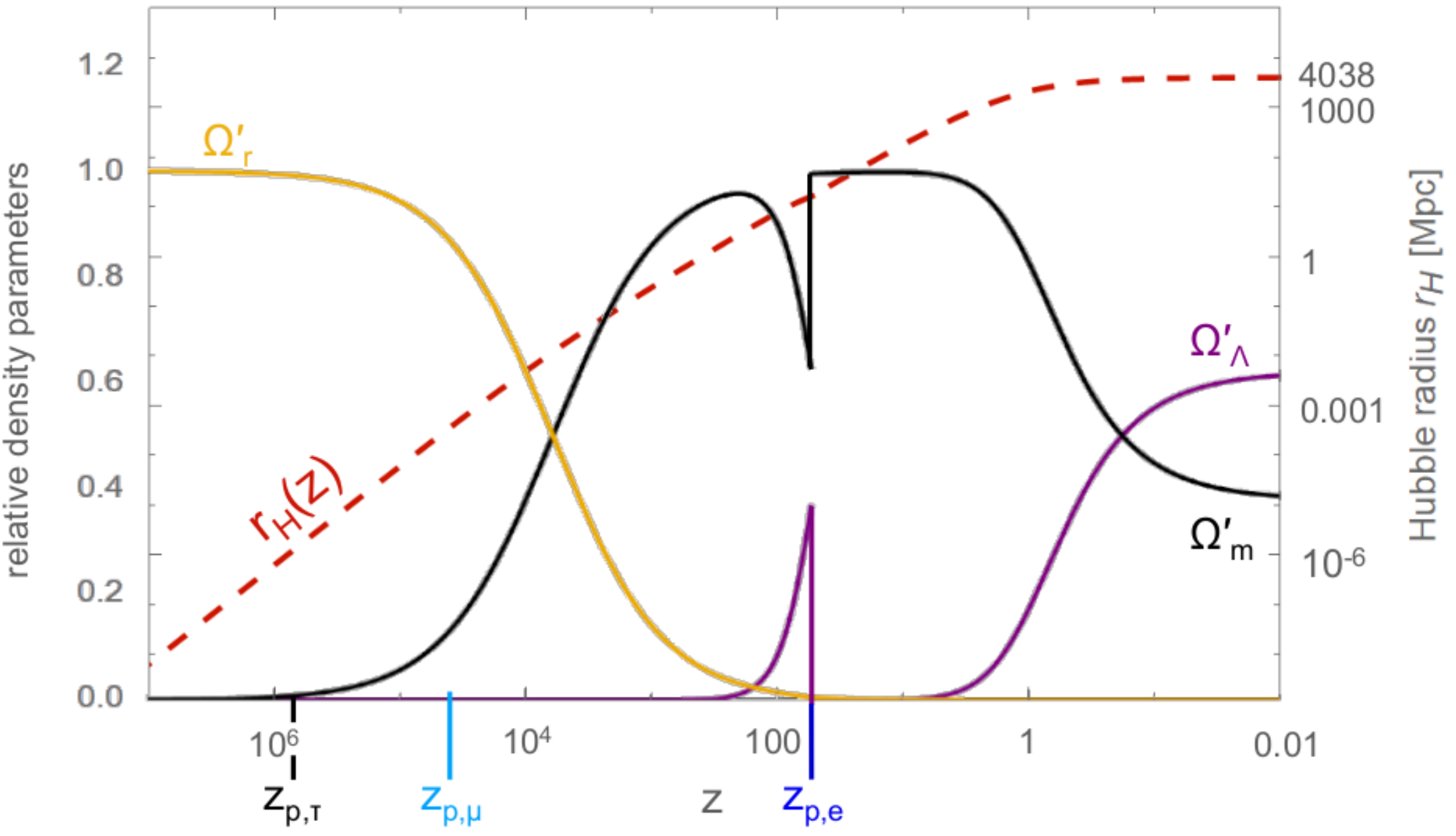}
\caption{Cosmological model SU(2)$_{\CMB}$ of \cite{Hahn:2018dih} (with parameter values fitted to the TT, TE, and EE CMB Planck power spectra and taken from column 2 of Table 2 of that paper) 
in terms of relative density parameters as functions of redshift $z$. Normalised density parameters refer to dark energy ($\Omega^\prime_\Lambda$), to total matter (baryonic and dark, $\Omega^\prime_{m}$), and to radiation (three flavours of massless neutrinos and eight relativistic polarisations in a CMB subject to SU(2)$_{\rm CMB}$, $\Omega^\prime_{r}$). The red dotted line represents the Hubble radius of this model. The redshifts of $e$-lump, $\mu$-lump, and $\tau$-lump depercolations are indicated by vertical lines intersecting the $z$-axis. Only e-lump depercolation is taken into account explicitly within the cosmological model SU(2)$_{\CMB}$ since at $z_{p,\mu}=40,000$ and $z_{p,\tau}=685,000$ the Universe is radiation dominated.}
\label{fig:rH}
\end{figure}

Next, we discuss question (ii). Consider a situation where the gravitational Bohr radius $r_B$ exceeds the Hubble radius $r_H(z)=H^{-1}(z)$ at some redshift $z$. Here $H(z)$ defines the Hubble parameter subject to a given cosmological model. In such a situation, the lump acts like a homogeneous energy density (dark energy) within the causally connected region of the Universe roughly spanned by $r_H$. If $r_B$ falls {\sl sizably} below $r_H$ then formerly homogeneous energy density may decay into isolated lumps. In order to predict at which redshift $z_p$ such a depercolation epoch has taken place we rely on the extraction of the epoch  $z_{p,e}=53$ in \cite{Hahn:2018dih} for the depercolation of e-lumps. To extract the depercolation redshifts $z_{p,\mu}$ and $z_{p,\tau}$ we use the cosmological model SU(2)$_{\rm CMB}$ proposed in \cite{Hahn:2018dih} with parameters values given in column 2 of Table 2 of that paper. In Fig.\,\ref{fig:rH} the relative density parameters of the cosmological model SU(2)$_{\rm CMB}$ are depicted as functions of $z$, and the point of e-lump depercolation $z_{p,e}=53$ is marked by the cusps in dark energy and matter. 

The strategy to extract $z_{p,\mu}$ and $z_{p,\tau}$ out of information collected at $z_{p,e}=53$ is to determine the ratio $\alpha_e$ of $r_H=16.4\,$Mpc at $z_{p,e}=53$ and $r_{B,e}=0.26$\,kpc for a typical, isolated, and unmerged e-lump as 
 \begin{equation}\label{ratiorHrB}
\left.\alpha_e\equiv\frac{r_H}{r_{B,e}}\right|_{z=z_{p,e}}=55,476\,. 
\end{equation}\nn
It is plausible that $\alpha_e$ can be promoted to a universal (that is, independent of the Yang-Mills scale and temperature) constant $\alpha$, again, because of the large hierarchy between all Yang-Mills scales to the Planck mass $M_P$. Moreover, the ratio of radiation temperature to the Planck mass $M_P$ remains very small within the regime of redshifts considered in typical CMB simulations. Using the cosmological model SU(2)$_{\rm CMB}$, Eq.\,(\ref{Bohrradiusfund}), and demanding $\alpha$ to set the condition for $\mu$- and $\tau$-lump depercolation ($r_H\equiv \alpha \, r_{B,i}$), one obtains  
 \begin{equation}
\label{zsperc}
z_{p,\mu}=\,40,000,\ \ \ \ z_{p,\tau}=685,000\,.
\end{equation}

In Fig.\,\ref{fig:rH} the relative density parameters $\Omega^\prime_\Lambda$ (dark energy), $\Omega^\prime_{m}$ for total matter (baryonic and dark), $\Omega^\prime_{r}$ (total radiation), and the Hubble radius $r_H$ are depicted as functions of $z$. Moreover, the redshifts of $e$-lump, $\mu$-lump, and $\tau$-lump depercolations -- $z_{p,e}$, $z_{p,\mu}$, and $z_{p,\tau}$  --  are indicated by vertical lines intersecting the $z$-axis. The depercolation epochs for $\mu$- and $\tau$-lumps at redshifts $z_{p,\mu}=40,000$, and $z_{p,\tau}=685,000$ are not modelled within SU(2)$_{\rm CMB}$ because the Universe then is radiation dominated.\nn

\begin{figure*}
\thisfloatpagestyle{empty}
\centering
\includegraphics[width=16.55cm]{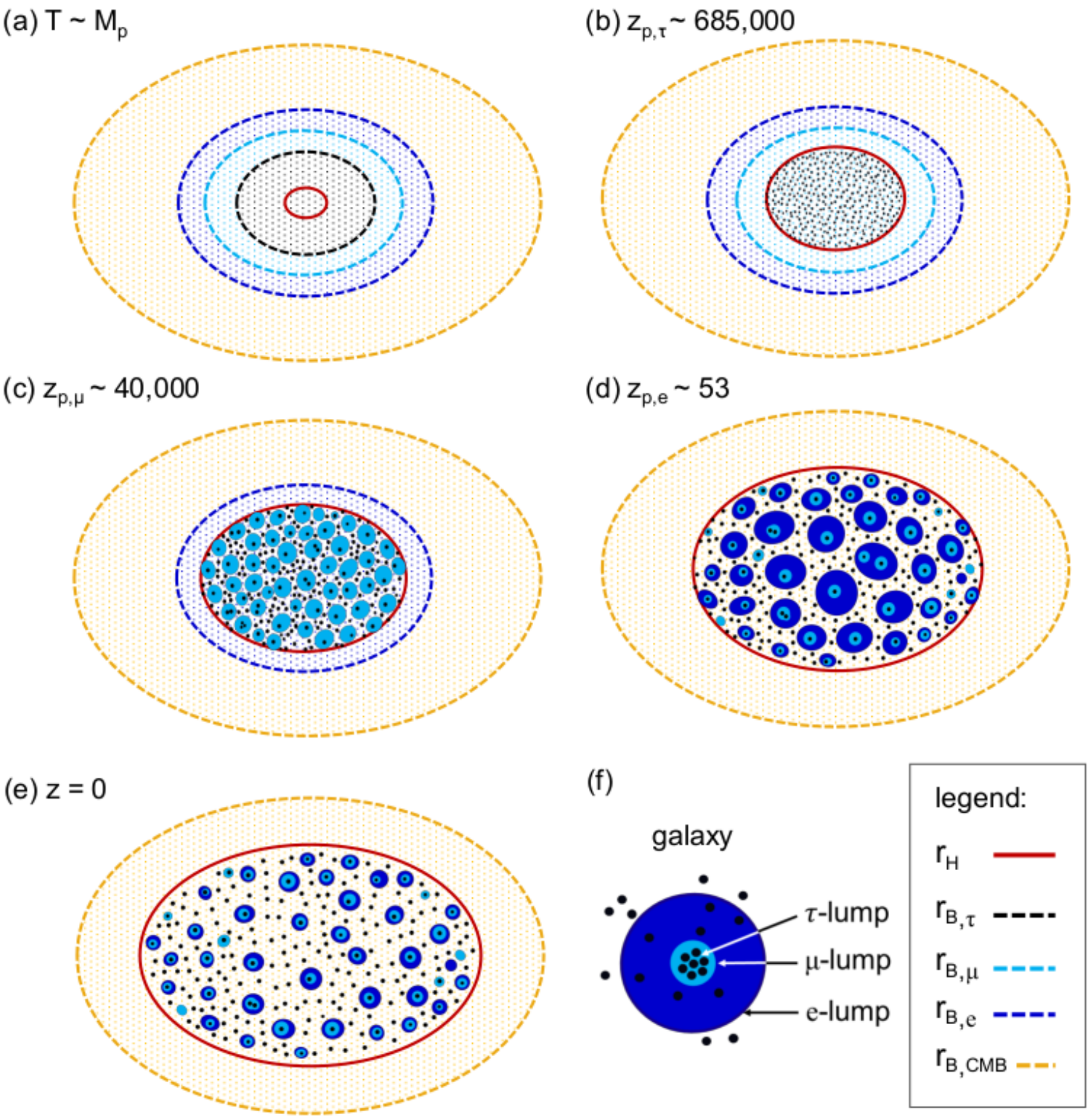}
\vspace{-4mm}\caption{
The evolution of the Universe's dark sector according to SU(2) Yang--Mills theories of scales $\Lambda_e=m_e/(15\times 118.6)$, $\Lambda_{\mu}=m_\mu/(15\times 118.6)$, $\Lambda_{\tau}=m_\tau/(15\times 118.6)$ (confining phases, screened), and $\Lambda_{\rm CMB}\sim 10^{-4}\,$eV (deconfining phase, unscreened) invoking Planck-scale induced axial anomalies. The horizon size, set by the Hubble radius $r_H$ at various epochs (\textbf{a} - \textbf{e}), is shown by a red circumference. At epoch (\textbf{a}) gravity induced chiral symmetry breaking at the Planck scale creates a would-be-Goldstone boson which, due to the axial anomaly, gives rise to four ultralight axionic particle species. Their gravitational Bohr radii $r_{B,\tau}$, $r_{B,\mu}$, $r_{B,e}$, and $r_{B,{\rm CMB}}$ are much larger than $r_H$. Therefore, the associated energy densities should be interpreted as dark energy. (\textbf{b}) As the radiation dominated Universe expands the smallest Bohr radius $r_{B,\tau}$ falls below $r_H$. Once the ratio $\alpha\equiv r_H/r_{B,\tau}$ is sufficiently large ($\alpha=$ 55,500) $\tau$-lumps depercolate ($z_{p,\tau}=$ 685,000). (\textbf{c}) As the Universe expands further the Bohr radius $r_{B,\mu}$ falls below $r_H$. When the ratio of $r_H$ and $r_{B,\mu}$ again equals about $\alpha=$ 55,500 $\mu$-lumps derpercolate ($z_{p,\mu}=$ 40,000). The cosmological matter densities of $\tau$ and $\mu$-lumps are comparable \cite{Hahn:2018dih}. Since the mass of an isolated, unmerged $\tau$-lump is by a factor of about $(m_\tau/m_\mu)^2\sim 283$ smaller than the mass of an isolated, unmerged $\mu$-lump it then follows that the number density of $\tau$-lumps is by this factor larger compared to the number density of $\mu$-lumps. (\textbf{d}) Upon continued expansion down to redshift $z_{p,e}=53$ e-lumps depercolate. Their number density is by a factor of $(m_\mu/m_e)^2\sim$ 42,750 smaller than the number density of $\mu$-lumps. (\textbf{e}) The value of $r_{B,{\rm CMB}}$ is vastly larger than $r_H(z=0)$: $r_{B,{\rm CMB}}=2.4\times 10^{10}\,$Mpc vs. $r_H(z=0)=4038\,$Mpc. Therefore, a depercolation of CMB-lumps up to the present is excluded. As a consequence, the condensate of CMB-axions {\sl is} dark energy. (\textbf{f}) Possible dark-matter configuration of a galaxy including $\tau$-lumps and a single $\mu$-lump inside an e-lump.
}
\label{fig:DarkHalosComic}
\end{figure*}

In Fig.\,\ref{fig:DarkHalosComic} a schematic evolution of the Universe's dark sector, subject to the SU(2) Yang-Mills theories SU(2)$_{\tau}$, SU(2)$_{\mu}$, SU(2)$_{e}$, and SU(2)$_{\rm CMB}$ invoking Planck-scale induced axial anomalies, is depicted. 

After a possible epoch of Planck-scale inflation and reheating the temperature of the radiation dominated Universe is close to the Planck mass $M_P$, and $r_H\sim M_P^{-1}$. In this situation, the Bohr radii of the various hypothetical lump species (Peccei-Quinn scale $M_P$, SU(2)$_{\tau}$, SU(2)$_{\mu}$, SU(2)$_{e}$, and SU(2)$_{\rm CMB}$ Yang-Mills dynamics) are  much larger than $r_H$, and the (marginal) dark sector of the model then solely contains dark energy. 
Around $z_{p,\tau}=685,000$ (radiation domination) the depercolation of $\tau$-lumps occurs for $\alpha\equiv r_H/r_{B,\tau}\sim 55,500$. Once released, they evolve like pressureless, non-relativistic particles and, cosmologically seen, represent dark matter.

%\vspace{-4mm}

%\thispagestyle{empty} 
%#######################################################################

%\newpage\thispagestyle{empty} 

As the Universe expands further, the ratio $\alpha\equiv r_H/r_{B,\mu}\sim 55,500$ is reached such that $\mu$-lumps start to depercolate at $z_{p,\mu}=40,000$. Since they contribute to the cosmological dark-matter density roughly the same amount like $\tau$-lumps, see \cite{Hahn:2018dih} for a fit of so-called primordial and emergent dark-matter densities to TT, TE, and EE power spectra of the 2015 Planck data, one concludes from Eq.\,(\ref{lumpmassratios}) that their number density is by a factor  $(m_\tau/m_\mu)^2\sim 283$ smaller than that of $\tau$-lumps. For a first estimate this assumes a neglect of local gravitational interactions. That is, at $\mu$-lump depercolation there are roughly 300 $\tau$-lumps inside one $\mu$-lump. Each of these $\tau$-lumps possesses a mass of $M_\tau=5.2\times 10^3\,M_\odot$. The implied accretion process involving additional $\tau$-lumps may catalyse the gravitational compaction of the thus contaminated $\mu$-lump, see discussion in Sec.\,5.1.\nn 

At $z_{p,e}=53$ e-lumps depercolate \cite{Hahn:2018dih}. Again, disregarding local 
gravitational binding, we conclude from Eq.\,(\ref{lumpmassratios}) and a nearly equal contribution of each lump species to the cosmological dark-matter density \cite{Hahn:2018dih} that the number densities of $\mu$- and $\tau$-lumps are by factors of $(m_\mu/m_e)^2\sim 42,750$ and $(m_\tau/m_e)^2\sim 283\times 42,750$, respectively, larger than the number density of e-lumps. At e-lump depercolation we thus have 42,750 $\mu$-lumps and $42,750\times 283\sim 1.2\times 10^7$ $\tau$-lumps within one e-lump.\nn

Again, ignoring local gravitational binding effects, the dilution of $\tau$- and $\mu$-lump densities by cosmological expansion predicts that today we have $42,750/(z_{p,e}+1)^3=0.27$ $\mu$-lumps and $42,750\times 283/(z_{p,e}+1)^3=77$ $\tau$-lumps within one e-lump. Local gravitational binding should correct these numbers to higher values but the orders of magnitude -- O(1) for $\mu$-lumps and O(100) for $\tau$-lumps -- should remain unaffected. 
It is conspicuous that the number of globular clusters within the Milky Way is in the hundreds \cite{GlobularClustersNumbers}, with typical masses between ten to several hundred thousand solar masses \cite{Ghez2008}. With $M_\tau=5.2\times 10^3\,M_\odot$ it is plausible that the 
dark-mass portion of these clusters is constituted by a single or a small number of merged 
$\tau$-lumps. In addition, in the Milky Way there is one central massive and dark object with about $(4.5\pm 0.4)\times 10^6$ \cite{Ghez2008} or $(4.31\pm 0.36)\times 10^6$ solar masses \cite{Gillessen:2008qv}. If, indeed, there is roughly one isolated $\mu$-lump per isolated $e$-lump today then the mass range of the Milky Way's dark-matter disk, interpreted as a merger of few isolated e-lumps, implies the mass range of Eq.\,(\ref{compobjMW}) for the associated $\mu$-lump merger. This range contains the mass of the central massive 
and dark object determined in \cite{Ghez2008,Gillessen:2008qv}.

\vspace{-1mm}

\section{Discussion, Summary, and Outlook}\label{Sec5}

\vspace{-1mm}

\subsection{Speculations on origins of Milky Way's structure}

The results of Sec.\,\ref{Sec5} on mass ranges of $\tau$-lumps, $\mu$-lumps, and e-lumps being compatible with typical masses of globular clusters, the mass of the central compact Galactic object \cite{Gillessen:2008qv,Ghez2008}, and the mass of the selfgravitating dark-matter disk of the Milky Way, respectively, is compelling. We expect that similar assignments can be made to according structures in other spiral galaxies. \nn

Could the origin of the central compact object in Milky Way be the result of $\tau$- and $\mu$-lump mergers?  As Fig.\,\ref{fig:SchwarzschieldradiusVSBohrradii} suggests, a merger of $n\ge 222$ isolated $\tau$- or $\mu$-lumps is required for black hole formation. Since we know that the mass of the central compact object is $\sim 4\times 10^6 M_\odot$ a merger of $n\ge 222$ $\mu$-lumps is excluded for Milky Way. Thus only a merger of $n\ge 222$ $\tau$-lumps, possibly catalysed by the consumption of a few $\mu$-lumps, is a viable candidate for black-hole formation in our Galaxy. Such a process -- merging of several hundred $\tau$-lumps within the gravitational field of a few merging $\mu$-lumps down to the point of gravitational collapse -- would be consistent with the results of \cite{Gillessen:2008qv,Ghez2008} who fit stellar orbits around the central massive object of Milky Way extremely well to a single-point-mass potential. Indeed, the gravitational Bohr radius of a $\mu$-lump is $7\times 10^{-6}\,$kpc while the closest approach of an S2 star to the gravitational center of the central massive object of Milky Way is $17\,{\rm lh} =  5.8 \times 10^{-7}\,$kpc \cite{Gillessen:2008qv}. Therefore, $\mu$-lumps need to collapse in order to be consistent with a point-mass potential.\nn  

The Milky Way's contamination with baryons, its comparably large dark-disk mass vs. the mass of the low-surface-brightness galaxies analysed in Sec.\,3, and possibly tidal shear from the dark ring and the dark halo during its evolution introduce deviations from the simple structure of a typical low-surface-brightness galaxy. Simulations, which take all the here-discussed components into account, could indicate how typical such structures are, rather independently of primordial density perturbations.\nn

Isolated $\tau$-, $\mu$-, and e-lumps, which did not accrete sufficiently many baryons to be directly visible, comprise dark-matter galaxies that are interspersed in between visible galaxies. The discovery of such dark galaxies, pinning down their merger-physics, and determinations of their substructure by gravitational microlensing and gravitational-wave astronomy could support the here-proposed scenario of active structure formation on sub-galactic scales.

\subsection{Summary and Outlook}

In this paper we propose that the dark Universe can be understood in terms of axial anomalies \cite{Adler:1969er,Bell:1969ts,Fujikawa:1979ay} which are invoked by screened Yang-Mills scales in association with the leptonic mass spectrum. This produces three ultra-light axion species. Such pseudo Nambu-Goldstone bosons are assumed to owe their very existence to a gravitationally induced chiral symmetry breaking with a universal Peccei-Quinn scale \cite{Peccei:1977ur} of order the Planck mass $M_P=1.22\times10^{19}\,$GeV \cite{Giacosa:2008rw}. We therefore refer to each of these particle species as {\sl Planck-scale} axions. Because of the relation $m_{a,i}=\Lambda_i^2/M_P$ the screened Yang-Mills scale $\Lambda_i$ derives from knowledge of the axion mass $m_{a,i}$.  Empirically, the here-extracted screened scale $\Lambda_e=287\,$eV points to the first lepton family, compare with \cite{Hofmann:2017lmu}. This enables predictions of typical lump and axion masses in association with two additional SU(2) Yang-Mills theories associating with $\mu$ and $\tau$ leptons.\nn

Even though the emergence of axion mass \cite{Peccei:1977ur} and the existence of lepton families \cite{Hofmann:2017lmu} are governed by the same SU(2) gauge principle, the interactions between these ultra-light pseudo scalars and visible leptonic matter is extremely feeble. Thus the here-proposed relation between visible and dark matter could demystify the dark Universe. An important aspect of Planck-scale axions is their Bose-Einstein, yet non-thermal, condensed state. A selfgravitating, isolated fuzzy condensate (lump) of a given axion species $i=e,\mu,\tau$ is chiefly characterised by the gravitational Bohr radius $r_{B,i}$ \cite{Sin1994} given in terms of the axion mass $m_{a,i}$ and the lump mass $M_i=M_{200,i}$ (virial mass), see Eq.\,\ref{Bohr}. As it turns out, for $i=e$ the information about the latter two parameters is contained in observable rotation curves of low-surface-brightness galaxies with similar extents. Realistic models for the dark-matter density profiles derive from ground-state solutions of the spherically symmetric Poisson-Schr\"odinger system at zero temperature and for a single axion species. These solutions describe selfgravitating fuzzy axion condensates, compare with \cite{Schive:2014dra}. Two such models, the Soliton-NFW and the Burkert model, were employed in our present extractions of $m_{a,e}$ and $M_e$ under the assumption that the dark-matter density in a typical low-surface brightness galaxy is dominated by a single axion species. Our result $m_{a,e}=0.675\times 10^{-23}\,$eV is consistent with the result of \cite{Bernal2017}: $m_{a,e}=0.554\times 10^{-23}\,$eV. Interestingly, such an axion mass is close to the result $10^{-25}\,$eV\,$\le m_a\le 10^{-24}\,$eV \cite{Hlo_ek_2018} obtained by treating axions as a classical ideal gas of non-relativistic particles -- in stark contrast to the Bose condensed state suggested by Eq.\,\ref{condfrBos} or the gas surrounding it with intrinsic correlations governed by large de-Broglie wavelengths. This value of the axion mass is considerably lower then typical lower bounds obtained in the literature: $m_a>2.9\times 10^{-21}\,$eV \cite{Nadler:2019hjw}, $m_a=2.5^{+3.6}_{-2.0}\times 10^{-21}\,$eV \cite{Maleki:2020sqn}, $m_a>3.8\times 10^{-21}\,$eV \cite{Irsic:2017yje}, and $m_a\sim 8\times 10^{-23}\,$eV in \cite{Schive:2014dra}.
We propose that this discrepancy could be due to the omission of the other two axion species with a mass spectrum given by Eqs.\,(\ref{axionmassemutau}). For example, the dark-matter and thus baryonic density variations along the line of sight probed by a  Lyman-$\alpha$ forest do not refer to gravitationally bound systems and therefore should be influenced by all {\sl three} axion species.\nn

Once axions and their lumps are categorised, questions about (i) the cosmological origin of lumps and (ii) their role in the evolution of galactic structure can be asked. Point (i) is addressed by consulting a cosmological model (SU(2)$_{\rm CMB}$ \cite{Hahn:2018dih}) which requires the emergence of dark matter by lump depercolation at defined redshifts, see also \cite{Hofmann:2020wvr}. 
Depercolation of e-lumps at redshift $z_{p,e} = 53$ anchors the depercolations of the two other 
lump species. One obtains $z_{p,\mu}=40,000$ and $z_{p,\tau}=685,000$. 

The critical temperature $T_{c,e}$ of SU(2)$_e$ for the deconfining-preconfining phase transition (roughly equal to the temperature of the Hagedorn transition to the confining phase \cite{bookHofmann}) is $T_{c,e}=9.49\,$keV \cite{Hofmann:2017lmu}. A question arises whether this transition could affect observable small-scale angular features of the CMB. In the SU(2)$_{\rm CMB}$ based cosmological model of \cite{Hahn:2018dih} $T_{c,e}=9.49\,$keV corresponds to a redshift of $z_{c,e}=6.4\times 10^7$. (Typically, CMB simulation are initialised at $z=10^9$ \cite{Ma_1995}). Traversing the preconfining-deconfining phase transition at $z_{c,e}$ an already strongly radiation dominated Universe receives additional radiation density and entropy. However, we expect that the horizon crossing of curvature perturbation at $z>z_{c,e}$, which may influence small-scale matter perturbations, will affect CMB anisotropies on angular scales $l>3000$ only. Therefore, Silk damping would reduce the magnitudes of these multipoles to below the observational errors.\nn

Up to the present, lump depercolation does not occur for the Planck-scale axion species associated with SU(2)$_{\rm CMB}$: Here the gravitational Bohr radius of the axion condensate always exceeds the Hubble radius by many orders of magnitude. As for point (ii), the masses and Bohr radii of $\mu$- and $\tau$-lumps seem to be related with the central massive compact object of the Milky Way \cite{Gillessen:2008qv,Ghez2008} and globular clusters \cite{Kalberla:2007sr}, respectively. Within a given galaxy such active components of structure formation possibly originate compact stellar streams through tidal forces acting on $\tau$-lumps. Whether this is supported by observation could be decided by a confrontation of N-body simulations (stars) in the selfgravitating background of the externally deformed lump.\nn  

Apart from cosmological and astrophysical observation, which should increasingly be able to judge the viability of the here-proposed scenario, there are alternative terrestrial experiments which can check the predictions of the underlying SU(2) gauge-theory pattern. Let us quote two examples: First, there is a predicted low-frequency spectral black-body anomaly at low temperatures ($T\sim 5\,$K) \cite{Hofmann:2009yh} which could be searched for with a relatively low instrumental effort. Second, an experimental link to SU(2)$_e$ would be the detection of the Hagedorn transition in a plasma at electron temperature $9.49\,$keV and the stabilisation of a macroscopically large plasma ball at a temperature of $1.3\times 9.49\,$keV \cite{Hofmann:2017lmu}. Such electron temperatures should be attainable by state-of-the-art nuclear-fusion experiments such as ITER or by fusion experiments with inertial plasma confinement.

\section{Data availability}
The SPARC library was analysed in support of this research \cite{Lelli:2016zqa}.
The processed data and program underlying this article will be shared on request to the corresponding author.

\end{multicols}{}

%\bibliography{BibTest2}

\bibliographystyle{ieeetr}
\bibliography{AxialAnomalies}

\end{document}